\documentclass[12pt]{article}

\pdfoutput=1
\usepackage[nosort]{cite}

\usepackage[pdftex]{graphicx}
\usepackage{epstopdf}

\usepackage{amsmath}
\usepackage{amsfonts}
\usepackage{amssymb}
\usepackage{multirow}
\usepackage{array}
\newcommand{\gsim}{\gtrsim}

\newcommand{\al}{\alpha}

\newcommand{\be}{\begin{equation}}
\newcommand{\ee}{\end{equation}}
\newcommand{\bee}{\begin{equation*}}
\newcommand{\eee}{\end{equation*}}
\newcommand{\bea}{\begin{eqnarray}}
\newcommand{\eea}{\end{eqnarray}}
\newcommand{\bean}{\begin{eqnarray*}}
\newcommand{\eean}{\end{eqnarray*}}

\begin{document}

\setcounter{page}{0}
\thispagestyle{empty}
%\pagestyle{empty}

%%%%%%%%%%%%%%%%%%%%%%%%%%%%%%%%%%%%%%%%%%%%%%%%%%%%%%%%%%%%%%%%%%%%%%%%%%%%%%%
\begin{flushright}
ANL-HEP-PR-09-90\\
CERN-PH-TH/2009-121\\
NU-HEP-TH/09-11\\
UCI-TR-2009-16
\end{flushright}

\vskip 26pt

\begin{center}
{\bf \LARGE {
Higgs in Space!
 }}
\end{center}

\vskip 24pt

\begin{center} {\large
{\bf C.~B. Jackson$^{a}$, G\'eraldine Servant$^{b}$, Gabe Shaughnessy$^{a,c}$,} \\[0.2cm]
{\bf Tim M.P. Tait$^{a,c,d}$ and Marco Taoso $^{b,e}$ } }
\end{center}

\vskip 20pt

\begin{center}

\centerline{$^{a}${\it
Argonne National Laboratory, Argonne, IL 60439, USA}}
\vskip .3pt
\centerline{$^{b}${\it CERN Physics Department, Theory Division, CH-1211 Geneva 23, Switzerland}}
\vskip .3pt
\centerline{$^{c}${\it Northwestern University, 2145 Sheridan Road, Evanston, IL 60208, USA}}
\vskip .3pt
\centerline{$^{d}${\it Department of Physics and Astronomy, University of California, Irvine, CA 92697. USA}}
\vskip .3pt
\centerline{$^{e}${\it IFIC (CSIC-Universitat de Val\`encia), Ed.Instituts, Apt. 22085, 46071 Valencia, Spain}}
\vskip 10.pt
\centerline{\tt  jackson@hep.anl.gov, geraldine.servant@cern.ch, ttait@uci.edu, }
\centerline{\tt g-shaughnessy@northwestern.edu, marco.taoso@ific.uv.es}
\end{center}
\vskip 13pt

\begin{abstract}
\vskip 3pt
\noindent
We consider the possibility that the Higgs can be produced in dark matter annihilations,
appearing as a line in the spectrum of gamma rays at an energy determined by the 
masses of the WIMP and the Higgs itself.  We argue that this phenomenon occurs
generally in models in which the the dark sector has large couplings to the 
most massive states of the SM and
provide a simple example inspired by the Randall-Sundrum vision of
dark matter, whose 4d dual corresponds to electroweak symmetry-breaking by strong dynamics
which respect global symmetries that guarantee a stable WIMP.
The dark matter is a Dirac fermion that couples to a $Z^\prime$ acting as a portal to the 
Standard Model through its strong coupling to top quarks.  Annihilation into light standard model degrees of freedom is suppressed and generates a feeble
continuum spectrum of gamma rays. Loops of top quarks mediate annihilation 
into $\gamma Z$, $\gamma h$, and $\gamma Z^\prime$, providing a 
forest of lines in the spectrum. Such models can be probed by the Fermi/GLAST 
satellite and ground-based Air Cherenkov telescopes.
\end{abstract}

\newpage

\tableofcontents

\vskip 13pt

%%%%%%%%%%%%%%%%%%%%%%%%%%%%%%%%%%%%%%%%%%%%%%%%%%%%%%%%%%%%%%%%%%%%%%%%%%%%%

\section{Introduction}
\label{sec:intro}

The cosmological evidence for dark matter (DM) is overwhelming, and yet we are still 
unsure of its nature.  We approach a new era of searching for dark matter, with the current generation of experiments closing in on weak scale masses and couplings.  Among 
the many alternate possible visions of dark matter, Weakly-Interacting Massive Particles 
(WIMPs) are among the most appealing, largely because of the 
potential to understand their population as a thermal relic, and the prospect of a
connection to the dynamics of electroweak symmetry breaking (EWSB) \cite{reviews}.

If WIMPs are indeed part of the dynamics of EWSB, they are expected to have relevant 
interactions with the fields of the Standard Model.  The next few years are likely to be very rewarding, as WIMPs may be produced in high energy particle accelerators, scattering non-relativistically with heavy nuclei, and/or annihilating into observable particles in space.  Each process represents a unique opportunity to learn about the nature of dark matter.  In particular, indirect detection of two WIMPs annihilating in our galaxy into high energy gamma
rays could potentially be our first indication that WIMPs are electroweakly active particles, 
and our first non-gravitational glimpse of dark matter.

WIMPs are dark, which implies that they do not couple directly to photons.  The 
processes by which observable gamma rays are produced are thus typically complicated.  
In most models, the dominant annihilation is into charged particles which can themselves 
radiate photons, hadronic states including $\pi^0$'s which decay into $\gamma \gamma$, 
and/or heavier states which decay into quarks and leptons.  The continuum of photons 
thus produced receives some imprint of the WIMP and its annihilation channels, but 
strong features are often lacking.  More striking are the (typically) subdominant
$2 \rightarrow 2$ reactions in which a WIMP annihilates directly into a photon and 
another particle, $X$.  The kinematics produce a line in the photon energy,
\bea
E_{\gamma} = M \left( 1 - \frac{M_X^2}{4 M^2} \right) 
\label{eq:energy}
\eea
with $M$ and $M_X$ respectively the WIMP and $X$ masses, which is a striking feature 
compared to expected astrophysical backgrounds. $\gamma$ line signals could also be produced by decaying dark matter, see {\it e.g.} \cite{Bertone:2007aw}. In this case, Eq.~\ref{eq:energy} is replaced by 
$E_{\gamma} = \frac{M}{2} ( 1 - \frac{M_X^2}{ M^2} ) $.
Such emission could be detected by current 
gamma-ray telescopes, such as Fermi-LAT, CANGAROO, HESS, MAGIC, and VERITAS,
which cover energies in the range of GeV - 10 TeV 
with resolutions of order $\Delta E / E \sim 0.1$.
Depending 
on the WIMP mass and couplings, several particles  may play the role of $X$, 
thus an entire forest of lines may be produced by WIMP annihilations \cite{Bertone:2009cb}, 
and their positions and intensities may reveal many features of the underlying theory of 
dark matter, including the presence of heavy states no longer present in the
Universe today.

It was recognized early on that supersymmetric dark matter may produce $\gamma Z$ 
as well as $\gamma \gamma$ final states \cite{Bergstrom:1997fh}.  However, the finite 
energy resolution of the experiments makes identifying the $\gamma Z$ line distinctly 
from the $\gamma \gamma$ line very challenging unless the WIMP is very
 light, like in the Inert Higgs DM Model \cite{Gustafsson:2007pc}.  In 
theories with more massive states, such as the $B^{(1,1)}$ Kaluza-Klein modes 
present in the 6d chiral square \cite{Dobrescu:2004zi} model of universal extra 
dimensions \cite{Appelquist:2000nn}, the process $B_H B_H \rightarrow \gamma B^{(1,1)}$ 
may reveal the presence of $B^{(1,1)}$ and help distinguish the chiral square from the 
5d case \cite{Servant:2002aq}.  Theories with light $Z^\prime$s that couple to WIMPs 
may also have a strong line from annihilation into $\gamma Z$ \cite{Dudas:2009uq}.

An interesting possibility is when $X$ is the Higgs boson $h$.  The apparent coincidence 
between the relic density and the electroweak scale raises the possibility that the 
dynamics of dark matter is related to that of the electroweak symmetry-breaking, and
such a connection suggests that WIMPs may have important couplings to massive states, 
such as top quarks, electroweak bosons, and the Higgs.  The observation of a line in 
the gamma ray spectrum whose position reflects the Higgs mass would be an exciting 
discovery, and could even constitute the first observation of a Higgs production 
process. However, 
it would still require observation and
measurement of the Higgs at colliders to decisively identify the 
particle associated with that line as the Higgs.  As Tevatron and LHC experiments strive 
to produce the Higgs in particle collisions, it is fascinating that dark matter annihilations 
may already be producing it in space.  Identification of a Higgs line would itself tell us 
much about the theory of WIMPs.  It would give credence to the notion that 
WIMPs are part of the dynamics of electroweak symmetry-breaking, and suggest 
properties such as the WIMP spin.  Scalar WIMPs carry no intrinsic angular momentum, 
and Majorana statistics allow $s$-wave annihilations only through a spin-singlet state, 
either of which lead to suppression of WIMP annihilations into $\gamma h$ by the tiny 
WIMP velocities $v$ expected in the galaxy.  
In principle, vector dark matter  could lead to a  $\gamma h$ signal. 
However, in popular theories such as Little Higgs models \cite{Perelstein:2006bq}
and Kaluza-Klein dark matter \cite{unpublished}
annihilation into  $\gamma h$ results from box diagrams and is highly suppressed. 
The observation of a $\gamma h$ line 
would thus favor particular combinations of WIMP spins and interactions.

In this article, we explore the possibility that WIMPs may annihilate into $\gamma h$, 
producing Higgs bosons in space in association with a photon whose energy reflects 
the mass of the Higgs (and the WIMP).  Our calculations are performed in the context 
of an effective theory describing  a Dirac fermion dark matter candidate.
Such a model captures the low energy physics of the Randall-Sundrum (RS) 
model of a warped extra dimension \cite{Randall:1999ee}, in which the WIMP 
is a right-handed neutrino whose stability results from the need to suppress rapid
proton decay \cite{Agashe:2004ci}.  However, we emphasize that this model is not unique -- 
as we argue in Section~\ref{sec:dmtop}, a WIMP whose dynamics is intimately linked 
with EWSB (and has appropriate statistics) is {\em likely} to annihilate in this way.  In 
Section~\ref{sec:gammaspectrum} we present the continuum of gamma rays expected when 
WIMPs couple strongly to top quarks, and  we compute
the expected intensity of the $\gamma Z$, $\gamma h$, and $\gamma Z^\prime$ lines
and discuss prospects for detection.  
We briefly discuss predictions for the relic density, direct detection, and some collider signals in Section ~\ref{sec:othersignals}. In Section~\ref{sec:conclusions}, we finish with an outlook and conclusions.

%================================================
\section{A Dark Matter--Top Quark Connection}
\label{sec:dmtop}

If dark matter arises as part of the dynamics of electroweak symmetry-breaking, it is 
natural to expect the WIMP to have couplings which favor the most massive states of the 
Standard Model.   
Here, we explore the possibility that the WIMP has important couplings to the top quark, 
through which it can couple at the loop level both to photons and to Higgs bosons.  

\subsection{Effective Theory}

We take the WIMP to be a Dirac fermion $\nu$ 
which is a singlet under the SM gauge interactions.
It is charged under a (spontaneously broken) $U(1)^\prime$ gauge symmetry, the massive
gauge boson of which acts as a portal to the SM by coupling to the top quark.  The effective Lagrangian
contains,
\bea
{\cal L}  & = &  {\cal L}_{SM}   - \frac{1}{4} \hat{F}^\prime_{\mu \nu} \hat{F}^{\prime \mu \nu} 
+ M_{\hat{Z}^\prime}^2 \hat{Z}^\prime_{\mu} \hat{Z}^{\prime \mu} 
+ \frac{\chi}{2} \hat{F}^\prime_{\mu \nu} \hat{F}_Y^{\mu \nu} 
+ \hat{g}_t^{Z^\prime} ~ \bar{t} \gamma^\mu P_R \hat{Z}^{\prime}_{\mu}  t 
\nonumber \\ & & ~~~~
+ i \bar{\nu} \gamma^\mu \left( \partial_{\mu} - i \hat{g}_\nu^{Z^\prime} P_R \hat{Z}^{\prime \mu}  \right) \nu
+ M_\nu \bar{\nu} \nu
\label{eq:langrangian}
\eea
where $\hat{F}^{\prime}_{\mu \nu}$ ($\hat{F}^{Y}_{\mu \nu}$) is the usual Abelian field strength for the $\hat{Z}^\prime$ (hypercharge boson), 
$\hat{g}_t^{Z^\prime}$ is the $\hat{Z}^\prime$ coupling to right-handed 
top quarks, and $\hat{g}_\nu^{Z^\prime}$ is the coupling to right-handed WIMPs.
$M_\nu$ is the WIMP mass.
One can easily include a coupling to the left-handed top (and bottom).
Our choice to ignore such a coupling fits well with typical RS models, balancing the need
for a large top Yukawa interaction with control over corrections to precision electroweak
observables.
The parameter $\chi$ encapsulates the strength of kinetic
mixing between the $Z^\prime$ and SM hypercharge bosons, and the hatted (unhatted)
quantities are those before (after) mixing, as discussed below.

We have included 
hypercharge-$\hat{Z}^\prime$ kinetic mixing through the term proportional to $\chi$.  
Such a term is consistent with the gauge symmetries, and even if 
absent in the UV,
will be generated in the IR description by loops of 
top quarks\footnote{$\chi$ can be engineered to vanish in the UV,
for example, by embedding $U(1)^\prime$ into a larger
gauge group which breaks down at scales of order $M_{\hat{Z}^\prime}$.}.   
The kinetic mixing parameter $\chi$ generates an effective coupling of SM states to
the $\hat{Z}^\prime$, and through electroweak symmetry breaking, mass mixing of the
$\hat{Z}^\prime$ with the SM $Z$ gauge boson resulting in 
a coupling of $\nu$ to the SM $Z$ boson. 
In terms of the quantities,
\bea
\eta &\equiv& \frac{\chi}{\sqrt{1-\chi^2}} \\
\Delta_Z & \equiv & \frac{M_{\hat{Z}^\prime}^2}{M^2_{Z_0}} \\ 
 M^2_{Z_0} & \equiv & \frac{1}{4} \left(g^2+g'^2 \right) v^2
\eea
the mass eigenvalues of the $Z$ and $Z^\prime$  
(the photon remains massless) are \cite{Kumar:2006gm},
\bea
M^2_{Z}&=& \frac{M_{Z_0}^2}{2}\left[1+s_W^2\eta^2+\Delta_Z \pm \sqrt{(1-s_W^2\eta^2-\Delta_Z)^2+4s_W^2\eta^2}\right]\\
M^2_{Z^\prime}&=& \frac{M_{Z_0}^2}{2}\left[1+s_W^2\eta^2+\Delta_Z \mp \sqrt{(1-s_W^2\eta^2-\Delta_Z)^2+4s_W^2\eta^2}\right]\\
\eea
where the upper  (lower) signs are for 
$\Delta_Z <1-s_W^2\eta^2$ ($\Delta_Z >1-s_W^2\eta^2$), 
with $s_W$ the sine of the weak mixing angle. 
For consistency with precision data, we require 
$\eta \ll 1$ for which $M_{Z} \approx M_{Z_0}$ and $M_{Z^\prime} \approx M_{\hat{Z}^\prime}$.

The relevant couplings are given by:
\bea
g_{\nu}^{Z}=(\hat{g}^{\nu}_R P_R+\hat{g}^{\nu}_L P_L) \  s_{\alpha} \ \sqrt{1+\eta^2}
\eea
\bea
g_{\nu}^{Z^{\prime}}=(\hat{g}^{\nu}_R P_R+\hat{g}^{\nu}_L P_L) \  c_{\alpha} \ \sqrt{1+\eta^2}
\eea
\bea
g_{\psi_{SM}}^{Z^{\prime}}=-\frac{g}{c_W}c_{\alpha}(t_{\alpha}+\eta s_W)\left[T_{3L} P_L-s_W^2Q\frac{t_{\alpha}+\eta /s_W}{t_{\alpha}+s_W\eta}\right]
\eea
\bea
g_{\psi_{SM}}^{Z}=-\frac{g}{c_W}c_{\alpha}(1-t_{\alpha}\eta s_W)\left[T_{3L} P_L-s_W^2Q\frac{1-t_{\alpha}\eta /s_W}{1-t_{\alpha}s_W\eta}\right]
\eea
with,
\bea
t_{\alpha} & = &
\frac{-2\eta s_W}{1-s^2_W \eta^2-\Delta_Z \pm \sqrt{(1-s^2_W\eta^2-\Delta_Z)^2+4\eta^2 s^2_W}}.
\eea

The parameters defining the model thus include the WIMP and $\hat{Z}^\prime$ 
masses  $M_{\nu}$ and $M_{\hat{Z}^\prime}\approx M_{Z^\prime}$,
the mixing parameter $\eta \approx \chi$, and the $Z^\prime$ 
couplings to $\nu$ and to right-handed
top quarks,  $g_{\nu}^{Z^\prime}\approx \hat{g}_{\nu}^{Z^\prime}$ and ${g}_{t}^{Z^\prime} \approx \hat{g}_{t}^{Z^\prime}$.
In addition we inherit the Higgs mass $m_h$ as an unknown parameter of the 
Standard Model itself.
Note that all the phenomenology of the model is controlled by the couplings of $Z^\prime$ to dark matter and to the top, and not through a direct coupling of dark matter to the Higgs \cite{Gopalakrishna:2009yz}.

As written, the model contains all of the most important features to describe dark matter, but
is not UV complete.  
It contains $U(1)^{\prime3}$ and mixed $U(1)^\prime$-SM
gauge anomalies which need to be cancelled, and the top Yukawa coupling
is not $U(1)^\prime$ invariant.  
One can imagine a variety of possible UV completions,
usually from additional massive fermions, whose presence is not expected to significantly 
affect the phenomenology of interest here (see Appendix \ref{sec:uvcompletion}
for one particular example).  
Such fermions will also contribute to $\chi$, and
motivate our treatment of it as an independent parameter.  
We have also neglected the Higgs sector which breaks the $U(1)^\prime$ symmetry,
giving mass to the $\hat{Z}^\prime$, as it is largely unimportant for the phenomenology of interest here.  

As a result of its incompleteness, our theory is not renormalizable, and it is important
to understand which quantities can be reliably computed in perturbation theory, and which are
sensitive to the unspecified UV physics.  At one-loop, four interactions are of particular
importance to describe the production of gamma rays from WIMP annihilation, the effective
$Z^\prime$-$\gamma$-$h$, $Z^\prime$-$\gamma$-$Z$, and 
$Z^\prime$-$\gamma$-$Z^\prime$ vertices, all mediated through loops of top quarks, and
the effective $Z^\prime$-$b$-$\bar{b}$ vertex, mediated by a loop of top quarks and $W$ bosons.
Naive power counting (confirmed by our explicit computations, see 
Section~\ref{sec:lines}) 
combined with the demands of Lorentz and $SU(2) \times U(1)$ gauge invariance indicates
that the $Z^\prime$-$\gamma$-$h$, $Z^\prime$-$\gamma$-$Z$, and 
$Z^\prime$-$\gamma$-$Z^\prime$ vertices are all finite, whereas the $Z^\prime$-$b$-$\bar{b}$
vertex is $\log$ divergent, and thus sensitive to the details of the UV completion.  While its
precise value is thus ill-defined in our effective theory framework, we expect it to be of
order,
\bea
g_{Z^\prime b \bar{b}} & \sim & g_t^{Z^\prime} \frac{\alpha }{32 \pi s^2_W} 
\frac{m_t^2}{M_W^2}
\log \left( \frac{ \Lambda^2}{Q^2} \right)
\label{eq:gzpbb}
\eea
where $\alpha$ is the fine-structure constant, $M_W$ is the $W$ boson mass, 
$\Lambda$ is the scale of the UV completion of the theory\footnote{Given the
need to generate the large top Yukawa interaction from higher dimensional
operators at scale $\Lambda$, this scale cannot
be much larger than $M_{Z^\prime}$ itself without the effective theory breaking down.  The
log in Equation~(\ref{eq:gzpbb}) is thus at most of order a few.}, and
$Q^2$ is a typical momentum transfer in the process of interest.
Where relevant, we will use this estimate (with $\log \Lambda^2 / Q^2 = 1$)
as the actual value
of the $Z^\prime$-$b$-$\bar{b}$ coupling, below.

\subsection{Connection to RS Models}

Our setup has a natural connection to RS theories in which the SM lives in the 
bulk \cite{Agashe:2003zs},
and such theories provide a natural UV completion.  In RS
theories, the hierarchy between the Planck and electroweak scales is explained through
warping of an extra dimension, with the Higgs living on the IR boundary where the natural scale of physics is $\sim$~TeV.  As a result of the localized Higgs, the zero mode
of the right-handed top quark must also live close to the IR brane, in order to realize the
large top mass\footnote{The left-handed top is usually chosen to be further from
the IR brane, in order to mitigate 
constraints from precision electroweak tests \cite{Agashe:2003zs}.}.

As a consequence of the warping, all of the low level KK modes have wave functions whose
support is concentrated near the IR brane.  Thus, they inevitably couple strongly to $t_R$ and
the Higgs.  If the dark matter particle is also a KK mode, for example a field whose boundary
conditions forbid the appearance of a zero mass state, it will also couple strongly to the
other KK modes.  Taken all together, these features establish the main features of our
effective theory.  To connect more precisely to specific models, one should
identify the DM and $Z^\prime$ fields as KK modes of bulk fields with the right properties.

Through the RS/CFT correspondence \cite{ArkaniHamed:2000ds,Rattazzi:2000hs}, 
the extra-dimensional theory is thought to be dual to an approximately scale-invariant 
theory in which most of the Standard Model is fundamental, but with the WIMP, Higgs, 
and right-handed top largely composite.  
The Higgs couples strongly to composite fields, and the amount of admixture in
a given SM fermion determines its mass \cite{Contino:2003ve}.
In this picture, the
$Z^\prime$ is one of the higher resonances, built out of the same preons as the 
WIMP and $t_R$.  RS theories provide a very motivated picture of the UV physics, 
but more generically, in any theory (not necessarily with approximate scale invariance) containing composite WIMPs \cite{Nussinov:1985xr} and composite 
top quarks \cite{Lillie:2007hd} belonging to a common sector \cite{Poland:2008ev}, 
one would expect strong couplings between them as a residual of the strong force 
which binds them, and perhaps negligible coupling to the rest
of the Standard Model.

Some RS constructions automatically contain SM gauge singlet bulk fermions which can
be identified with $\nu$.  The most obvious exists in models with an $SO(10)$ GUT
symmetry in the bulk, for which the SM matter fits into $16$ representations, including
a gauge singlet with $(-,+)$ boundary conditions whose mass
is generally somewhat atypically low compared to the other KK modes \cite{Agashe:2004ci}.
In this model, dark matter is stable as a result of a global symmetry needed to protect
against too-rapid proton decay.  Other possibilities include cases
in which the dark matter is a stable neutral
component \cite{Haba:2009xu,Carena:2009yt}
of a doublet under $SU(2)_L \times SU(2)_R$ 
(introduced to control precision corrections to $\Delta T$ \cite{Agashe:2003zs}), though
such objects have full strength $SU(2) \times U(1)$ interactions which we have not considered in this work.

In the $SO(10)$ model, the $Z^\prime$ represents the lowest KK mode of the 
$U(1)$ contained in $SO(10)$.  It typically has mixing with the electroweak
bosons, resulting in strong constraints from precision data.  This will also be
the case when the $Z^\prime$ is a KK mode of the electroweak bosons.  As we will see,
heavy $Z^\prime$s still lead to interesting signals, and can satisfy relic density
and direct detection constraints, provided their couplings are strong enough.
We circumvent these constraints by considering a $Z^\prime$ whose mixing with the 
$Z$ is kinetic.  At large $Z^\prime$ masses this is not operationally different
from the mass-mixing case, but it allows us to consider lower mass $Z^\prime$s which
are not ruled out by precision data.

As a final comment, the $SO(10)$ model also contains an electrically charged 
color triplet vector boson $X_s^\mu$ which couples directly to top and $\nu$, 
and has mass of order the typical KK scale.  We have chosen not to 
include $X_s^\mu$ in our effective theory, because it never approaches resonant behavior 
in WIMP annihilations, and thus is usually subleading compared to the $\hat{Z}^\prime$.

%================================================
\section{Gamma Ray Spectrum from DM Annihilation}
\label{sec:gammaspectrum}
\subsection{Continuum Photon Emission}
\label{sec:continuum}

\begin{figure}[t]
\begin{center}
\includegraphics[width=0.4\textwidth]{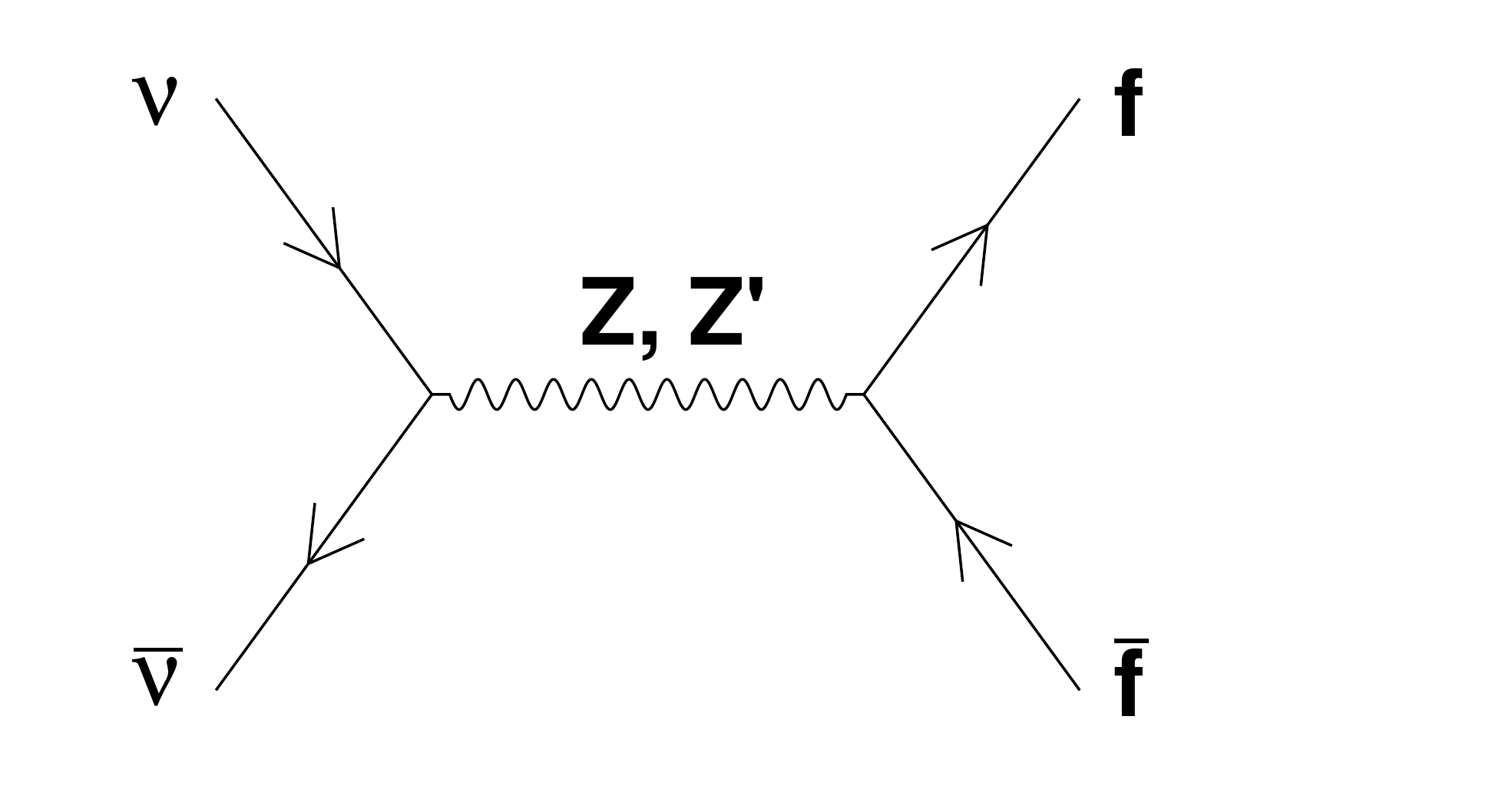} 
\includegraphics[width=0.4\textwidth]{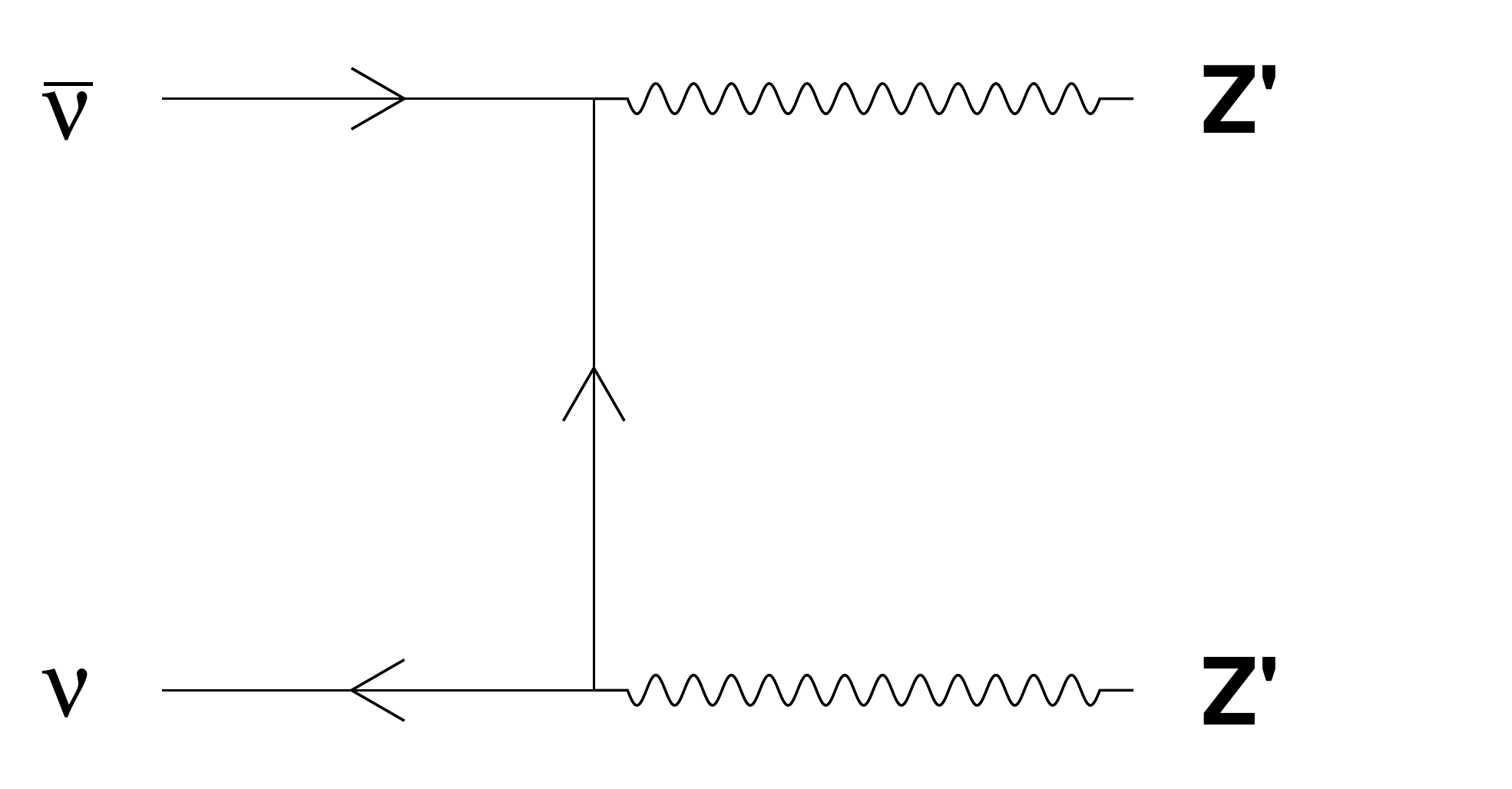} 
\includegraphics[width=0.4\textwidth]{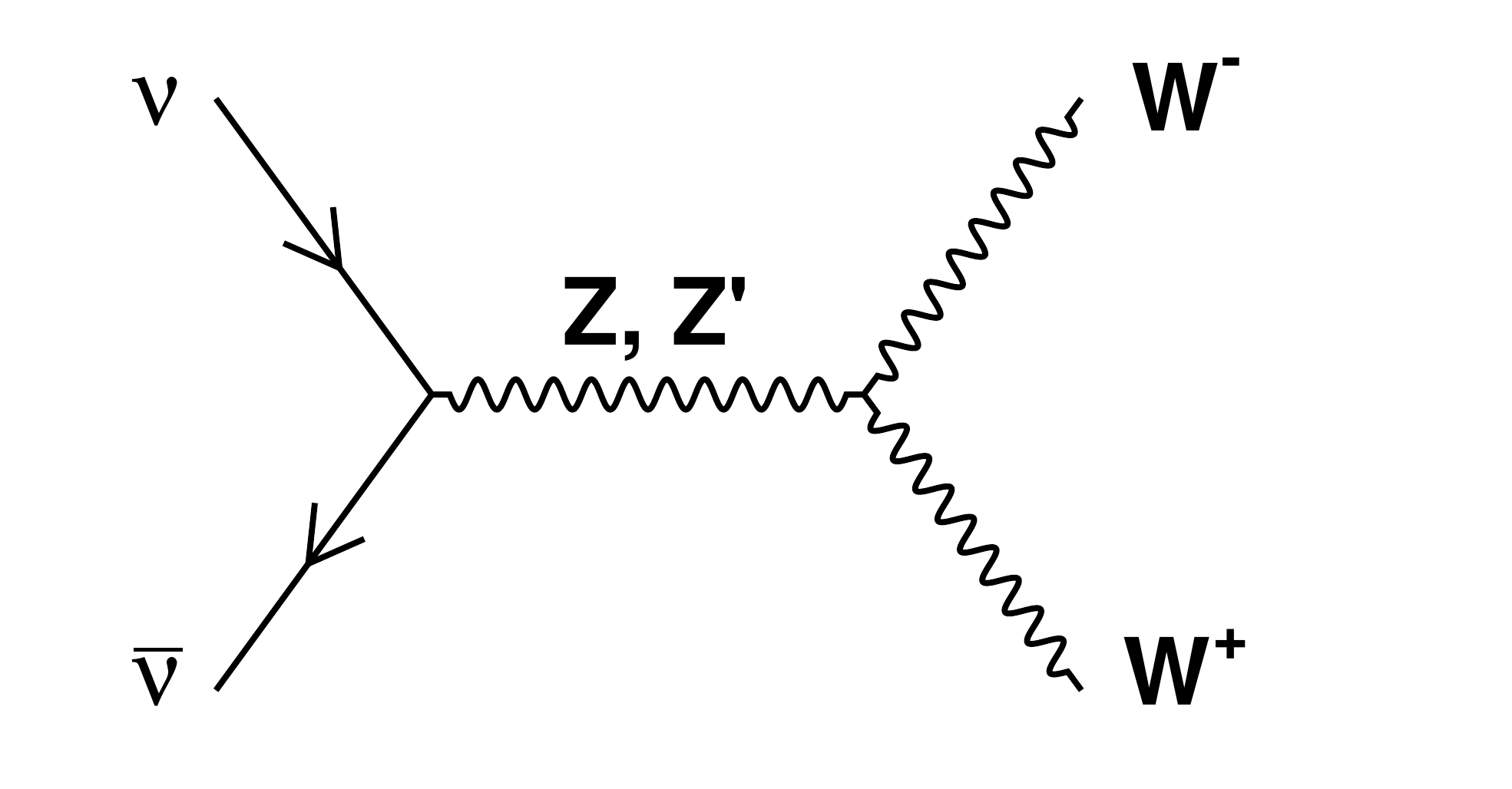} 
\includegraphics[width=0.4\textwidth]{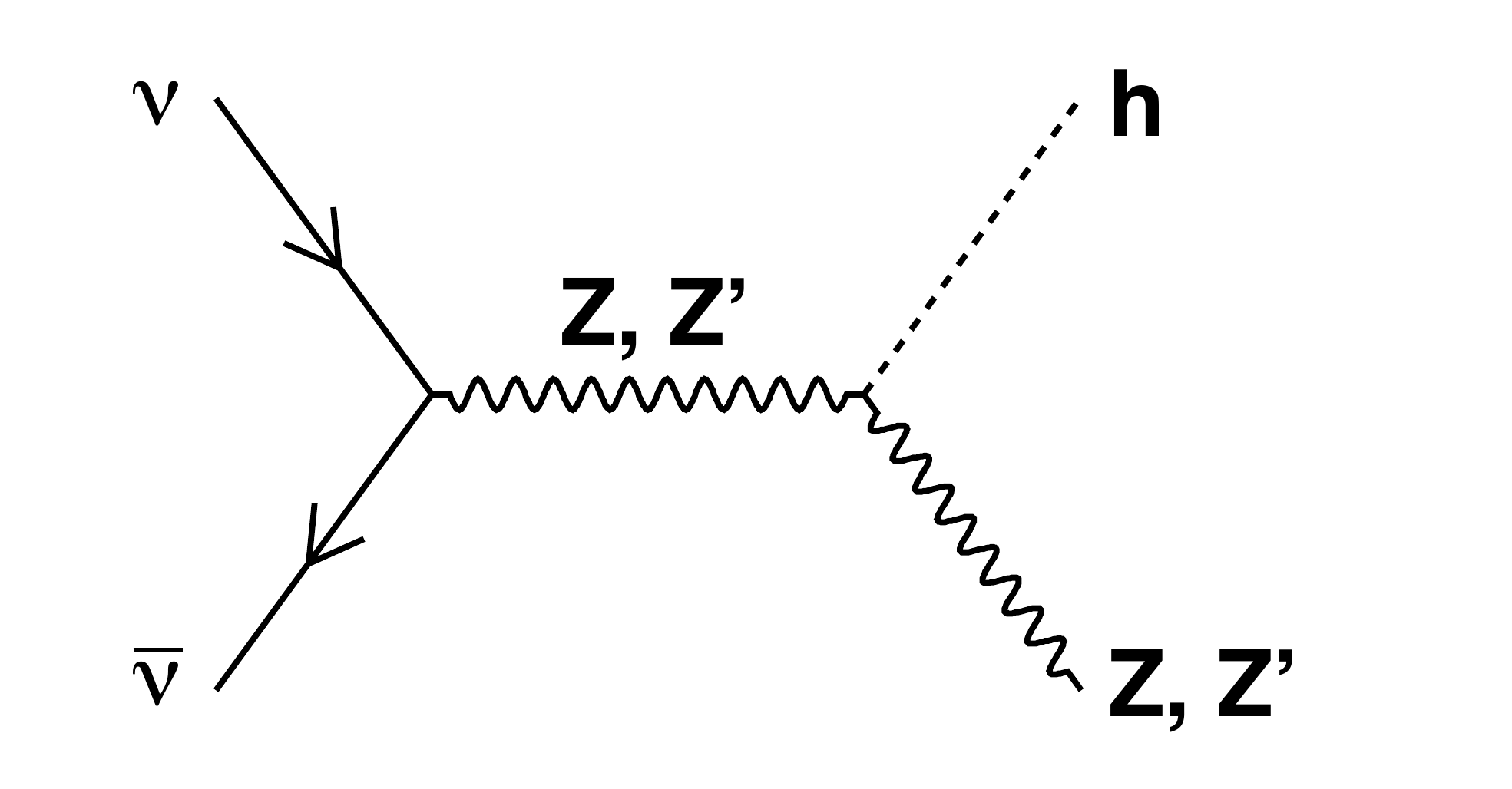} 
\caption[]{\small Representative tree-level Feynman diagrams for $\nu \bar{\nu}$ 
annihilation channels. If kinematically allowed, the two dominant channels are the $s$-channel $Z^{\prime}$ annihilation into $t \overline{t}$ and the $t$-channel annihilation into $Z^{\prime}$. All other processes are suppressed by the kinetic mixing $\eta$ or are loop-level.}
\label{fig:anni_diagrams}
\end{center}
\end{figure}

The continuum photon emission largely originates
from the decays of $\pi^0$ produced by hadronization
of strongly interacting states and from gamma-rays emitted
by light charged particles. 
As alluded to in the introduction, the resulting spectrum contains only very subtle information
about the primary products of dark matter annihilation, as their shapes are very similar.
In the particular case of $\nu \bar{\nu}$ annihilation, the dominant processes are those
shown in Fig.~\ref{fig:anni_diagrams}.

\begin{figure}[t]
\begin{center}
\includegraphics[width=0.485\textwidth]{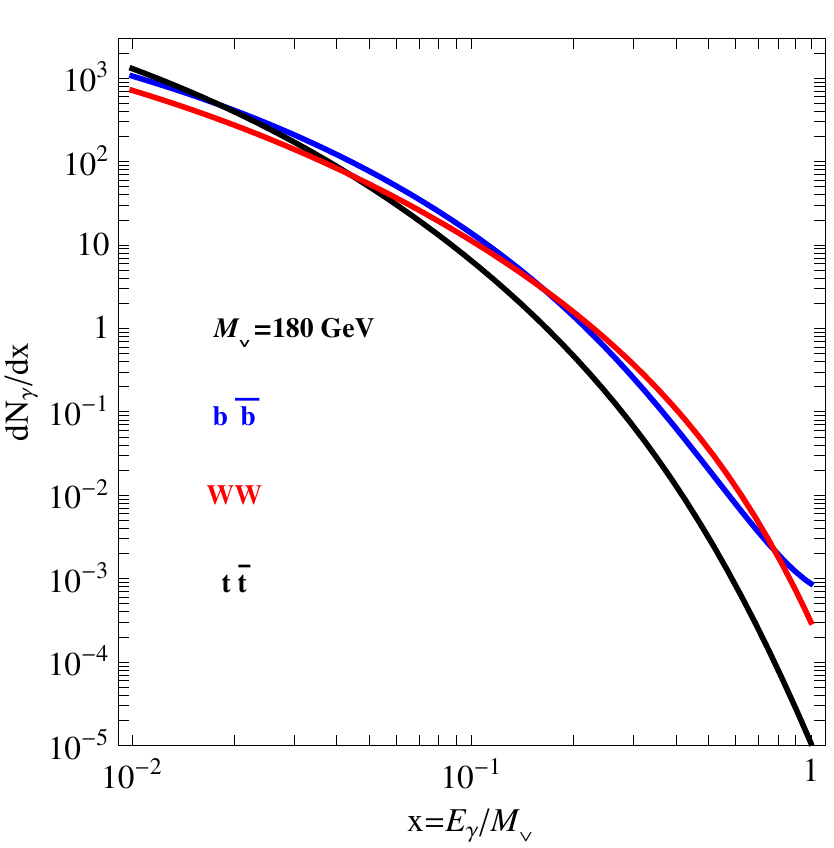} 
\includegraphics[width=0.485\textwidth]{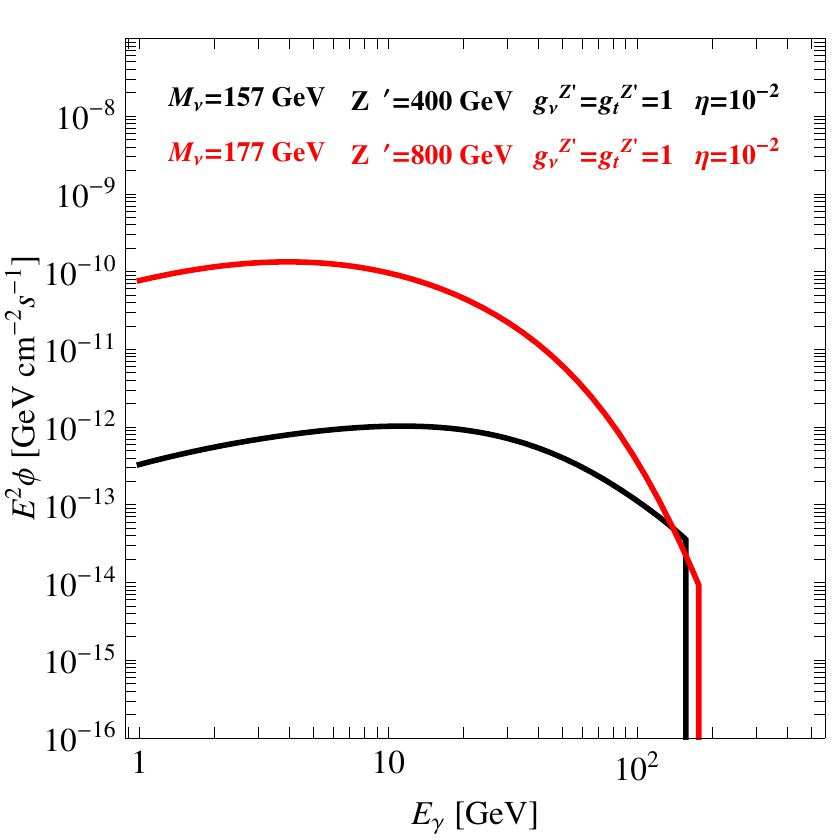} 
\caption[]{\small (Left panel) Continuum photon spectrum $dN_{\gamma}/dx,$ with $x=E_{\gamma}/M_{DM},$ obtained from annihilation into $b\overline{b}, WW$ or $ t\overline{t}$. 
(Right panel) Comparison of the continuum spectra obtained for two parameter sets with
$M_{\nu}<m_t$ and $M_{\nu}>m_t$. Fluxes are for an observation
of a $\Delta \Omega =10^{-5}$ sr region around the galactic center.
A NFW density profile has been employed. 
%For comparison, HESS~\cite{Aharonian:2009zk} and 
%EGRET~\cite{MayerHasselwander:1998hg} data are also shown.
}
\label{fig:continuum}
\end{center}
\end{figure}

One interesting aspect of our model is that it leads to suppressed continuum emission,
which increases the prominence of the gamma ray lines.
For $M_{\nu}$ less than both $m_t$ and $M_{Z^\prime}$, the natural 
annihilation into top pairs is closed, forcing annihilation predominantly 
into light SM particles, whose rates are suppressed by the small kinetic mixing $\eta$
or the loop-induced $Z^\prime$-$b$-$\bar{b}$ interaction.
For $M_{\nu}$ greater than either $m_t$ or $M_{Z^\prime}$, 
$\nu\bar{\nu}$ annihilation is dominantly into top quark pairs, and 
the continuum emission is more sizable. 
Nevertheless, since the continuum generated by $t\overline{t}$ annihilation is 
softer for $M_{DM} \lesssim 200$ GeV
than the one obtained from $b\overline{b}$ or $WW$ annihilation, 
even in this case, we expect reduced continuum emission close to the cut-off of
the spectrum, where the lines are located.
In Fig.~\ref{fig:continuum} we show the spectra for generic annihilation into 
$b\overline{b}, WW$ or $ t\overline{t}$ and spectra for $\nu \bar{\nu}$ annihilation for two parameter
sets with masses above and below $m_t$, both of which lead to the correct
thermal relic density (see Section~\ref{sec:relic}). The photon fluxes induced
by $\nu \bar{\nu}$ annihilations have been computed for a Navarro-Frenk-White DM density profile and
refer to an observation of the galactic central region
with an angular acceptance $\Delta \Omega =10^{-5}$ sr 
(see Section~\ref{sec:distribution} form more details).
%Also shown for comparison are the HESS 
%observations of the same angular region ~\cite{Aharonian:2009zk}
%and the EGRET data corresponding instead
%to about a $1^\circ$ region~\cite{MayerHasselwander:1998hg}.

%

%================================================
\subsection{Gamma Ray Lines}
\label{sec:lines}

In this section we compute the expected gamma ray line intensities, for the
$\gamma h$, $\gamma Z$ and $\gamma Z^\prime$ lines.  An interesting feature that 
results from an
$s$-channel $Z^\prime$ being the sole portal from the WIMP sector to the SM is the fact that there is no $\gamma \gamma$ line, as dictated by the
Landau-Yang theorem \cite{Yang:1950rg} (see also \cite{Keung:2008ve} for a more recent discussion). 
At leading order in the WIMP relative velocity, the cross section into $\gamma X$ is given by,
\bea
\label{eq:sigma-v}
\sigma v & = & \frac{1}{64 \pi M_\nu^2} \left( 1 - \frac{M_X^2}{4 M_\nu^2} \right) \overline{| {\cal M} |^2} \,.
\eea
In the sections below, we outline the computation of the the matrix elements ${\cal M}$ for annihilation into $\gamma h$, $\gamma Z$ and $\gamma Z^\prime$.

\subsubsection{$\gamma h$}

\begin{figure}[t]
\begin{center}
\includegraphics[width=0.45\textwidth]{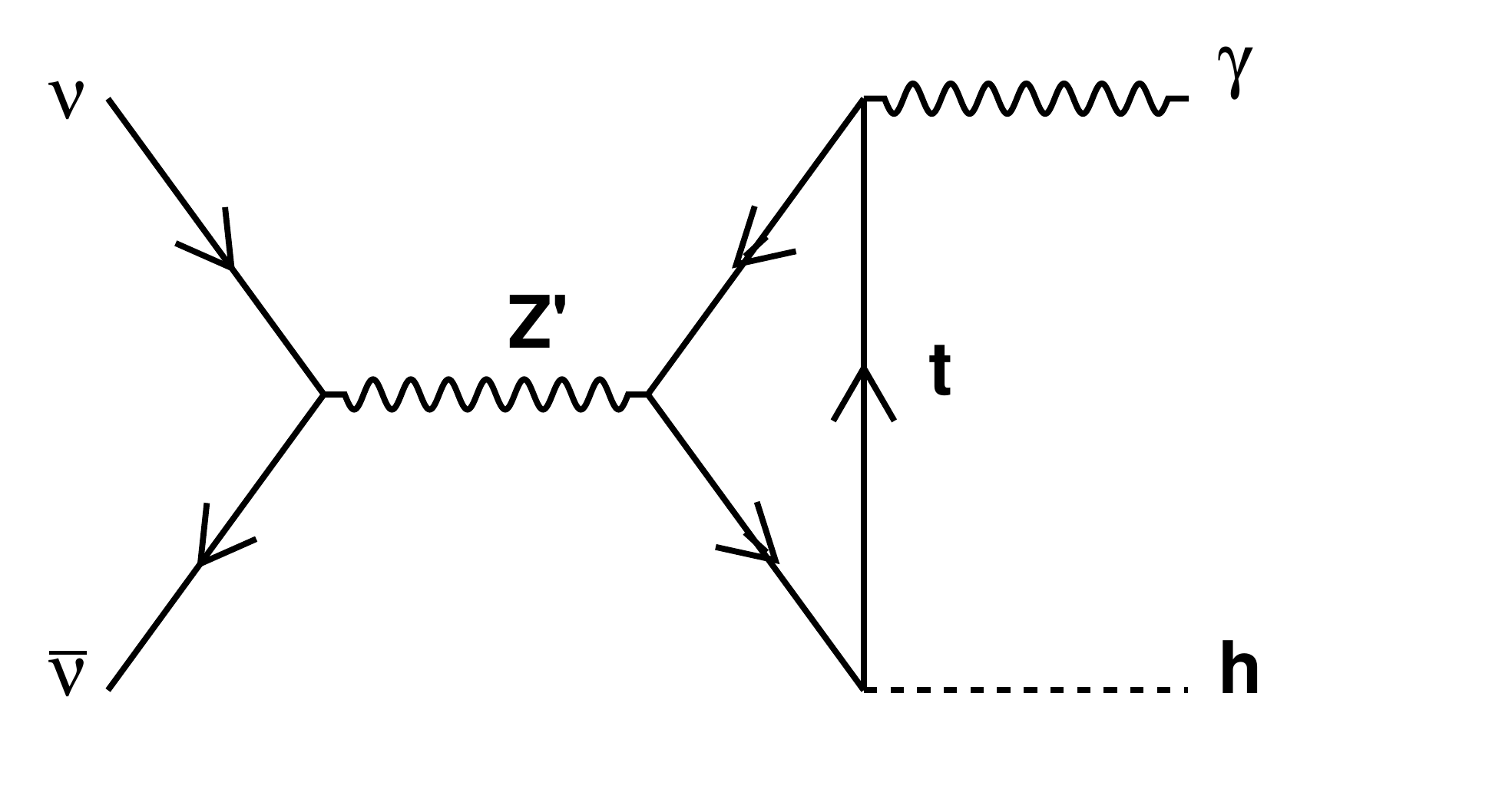} 
\end{center}
\caption[]{\small Representative Feynman diagram for $\nu \bar{\nu} \rightarrow \gamma h$.}
\label{fig:feyn1}
\end{figure}

A heavy neutrino and anti-neutrino can annihilate into a $\gamma h$ final state
through a $Z^\prime$ boson which connects to a loop of top quarks (see Fig.~\ref{fig:feyn1}).
In the non-relativistic limit, the matrix element, averaged over initial heavy neutrino spins and summed over the
final state photon polarizations may be expressed as,
\bea
\overline{|{\cal M}|^2}_{\gamma h}  & = & 
\frac{  \alpha \alpha_t N_c^2}{72 \pi^2} {\cal V}^2
\frac{ \left(g_t^{Z^\prime}  g_\nu^{Z^\prime} \right)^2 M_\nu^2 m_t^2}
{\left( 4 M_\nu^2 - M^2_{Z^\prime} \right)^2 +  M^2_{Z^\prime} \Gamma^2_{Z^\prime}}
\eea
where $N_c=3$ is the number of colors, $\alpha_t = y_t^2 / (4 \pi)$ is the fine structure constant
corresponding to the top Yukawa interaction, and $\Gamma_{Z^\prime}$ is the $Z^\prime$
width.

The vertex factor ${\cal V}$ describes the effective coupling between an off-shell $Z^\prime$ boson
and on-shell Higgs and photon, and may be expressed as,
\bea
{\cal V} _{\gamma h} & = &
\frac{8 M_\nu^2}{ \left( m_h^2 - 4 M_\nu^2 \right)} 
\left[ B_0 \left(m_h^2; m_t, m_t \right) - B_0 \left(4 M_\nu^2; m_t, m_t \right) \right]\nonumber\\
&+& \left[ 4 M_\nu^2 + 4 m_t^2 - m_h^2 \right] C_0 \left(m_h^2, 0, 4 M_\nu^2; m_t, m_t, m_t \right)
+ 2
\eea
where the scalar integrals are defined as,
\bea
\hspace*{-0.5cm}
\label{eq:B0}
B_0(p^2; m, m) & = & 16 \pi^2 \int \frac{d^n \ell}{(2 \pi)^n} \frac{1}{\ell^2 - m^2} \frac{1}{(\ell+p)^2 - m^2} ~, \\
C_0 \left(p_a^2, p_{b}^2, (p_a + p_b)^2; m_1, m_2, m_3 \right) 
& = & 16 \pi^2 \int \frac{d^n \ell}{(2 \pi)^n} 
\frac{1}{\ell^2 - m_1^2} \frac{1}{(\ell+p_a)^2 - m_2^2} 
\frac{1}{(\ell+p_a+p_b)^2 - m_3^2} ~.
\nonumber
\eea

\subsubsection{$\gamma Z$ and $\gamma Z^\prime$}

The calculation of the $\gamma Z$ and $\gamma Z^\prime$ cross sections follows 
along the same lines as that of $\gamma h$, but is somewhat more complicated by
the $Z$ (or $Z^\prime$) spin indices.  In this subsection, we outline the calculation of 
the cross sections for $\nu\bar{\nu} \to \gamma Z (Z^\prime)$ and reserve detailed
expressions for the appendix.

\begin{figure}[t]
\begin{center}
\includegraphics[width=0.35\textwidth]{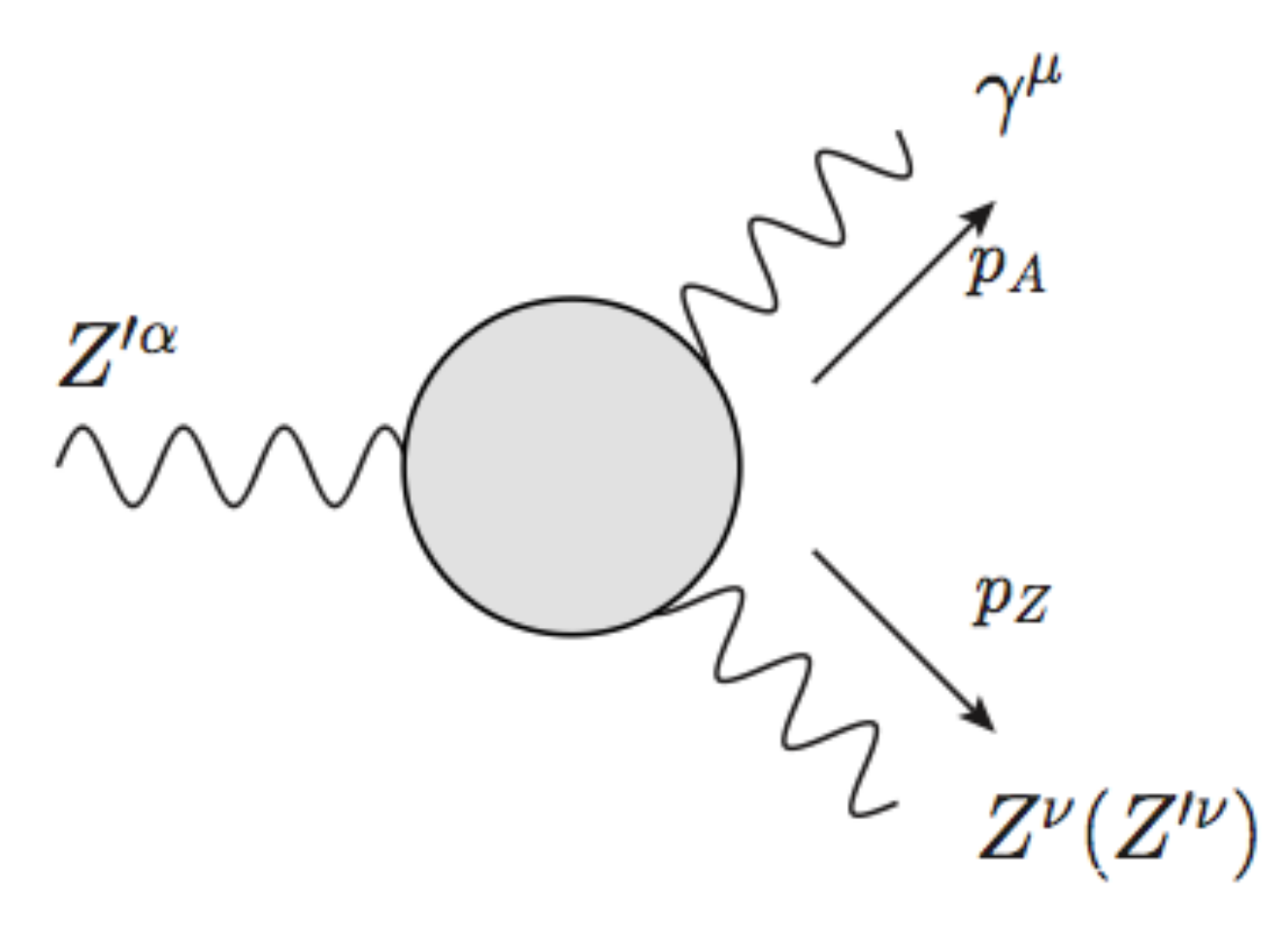} 
\end{center}
\caption[]{\small Labels for the  $Z^\prime$-$\gamma$- $Z / Z^\prime$ effective vertex.}
\label{fig:ZAZ-effvertex}
\end{figure}

We begin with the effective vertex depicted in Fig.~\ref{fig:ZAZ-effvertex}.  
The expression for the effective vertex is given by:
\begin{equation}
\Gamma^{\al} = \epsilon_\mu^*(p_A) \epsilon_\nu^*(p_Z)
  \sum_{i=1}^5 C_i M_i^{\al \mu \nu} \,.
\label{eq:ZgamZ_eff_vertex}
\end{equation}
where $p_A$ and $p_Z$ are the momenta of the photon and the outgoing 
$Z$ ($Z^\prime$), respectively,
and we have accounted for the transverality of the polarization tensors
($\epsilon^*(p_A) \cdot p_A = \epsilon^*(p_Z) \cdot p_Z = 0$).
The $C_i$ coefficients are functions of the top quark couplings to $Z$ and 
$Z^\prime$ as well as the scalar integrals given in Eqs.~(\ref{eq:B0}) 
with the replacement $m_h \to M_Z$.  The exact expressions for
these coefficients are given in appendix~\ref{app:ZA-coeffs}.  The $M_i^{\al \mu \nu}$ 
tensor structures
are expressed in terms of Levi-Civita tensors as:
\begin{eqnarray}
\label{eq:M1}
M_1^{\al \mu \nu} &=& p_{A,\lambda} \epsilon^{\lambda \mu \nu \alpha} \,,\\
\label{eq:M2}
M_2^{\al \mu \nu} &=& p_{Z,\lambda} \epsilon^{\lambda \mu \nu \alpha} \,,\\
\label{eq:M3}
M_3^{\al \mu \nu} &=& p_A^\alpha p_{A,\lambda} p_{Z,\sigma} 
  \epsilon^{\lambda \sigma \mu \nu} \,,\\
\label{eq:M4}
M_4^{\al \mu \nu} &=& p_Z^\alpha p_{A,\lambda} p_{Z,\sigma} 
  \epsilon^{\lambda \sigma \mu \nu} \,,\\
\label{eq:M5}
M_5^{\al \mu \nu} &=& p_A^\nu p_{A,\lambda} p_{Z,\sigma} 
  \epsilon^{\lambda \sigma \alpha \mu} \,.
\end{eqnarray}
After summing over the polarizations of the external gauge bosons, only 
$M_1$, $M_2$ and $M_5$ actually contribute to the matrix element squared.

Coupling the effective vertex $\Gamma^\alpha$ to the $Z^\prime \nu \bar{\nu}$ 
vertex and averaging (summing) over initial (final) state spins and polarizations,
we find the matrix-element-squared for $\nu\bar{\nu} \to \gamma Z$ is given by:
\begin{equation}
\overline{|{\cal M}|^2}_{\gamma Z} =  
\frac{\alpha^2 N_c^2}{576 \pi^2 s_W^2 c_W^2} {\cal V}_{\gamma Z}^2
\frac{ \left(g_\nu^{Z^\prime} \right)^2 M_\nu^2}
{\left( 4 M_\nu^2 - M^2_{Z^\prime} \right)^2 +  M^2_{Z^\prime} \Gamma^2_{Z^\prime}}\,,
\label{eq:Msq-ZA}
\end{equation}
where $s_W (c_W)$ is the sine (cosine) of the SM weak-mixing angle.
The matrix-element-squared for $\nu\bar{\nu} \to \gamma Z^\prime$ is:
\begin{equation}
\overline{|{\cal M}|^2}_{\gamma Z^\prime} =  
\frac{\alpha N_c^2}{144 \pi^3} {\cal V}_{\gamma Z^\prime}^2
\frac{ \left(g_\nu^{Z^\prime} \right)^2 M_\nu^2}
{\left( 4 M_\nu^2 - M^2_{Z^\prime} \right)^2 +  M^2_{Z^\prime} \Gamma^2_{Z^\prime}}\,.
\label{eq:Msq-ZpA}
\end{equation}

\begin{figure}[t]
\begin{center}
\includegraphics[width=0.49\textwidth]{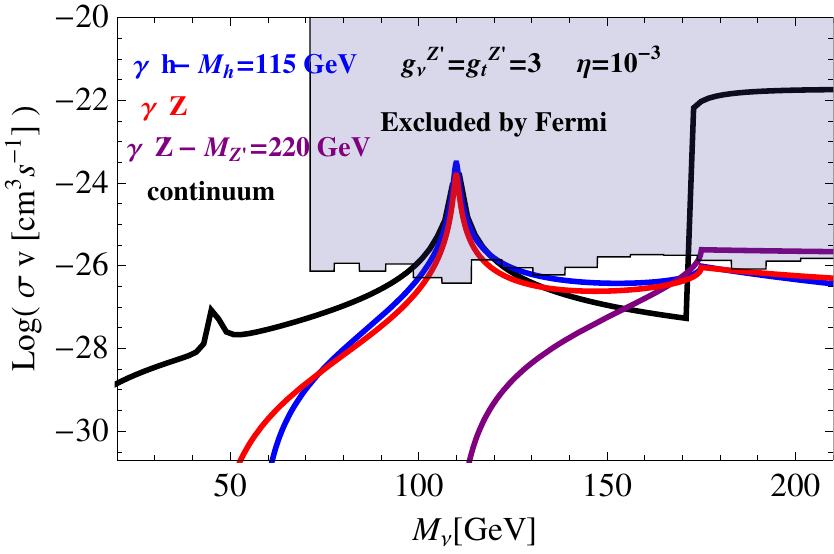} 
\includegraphics[width=0.49\textwidth]{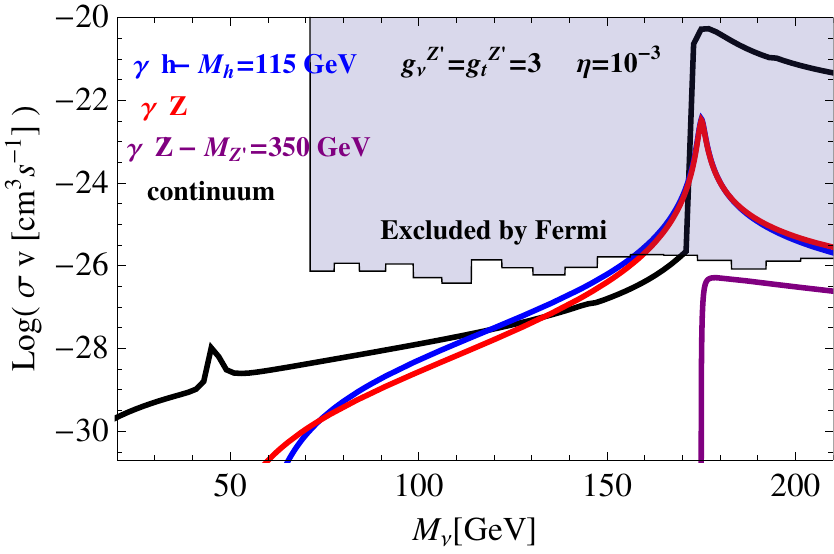} 
\caption[]{\small Cross sections as a function of the WIMP mass  for two values
of $M_{Z^\prime}$ and some choice of couplings for the three gamma line emission  channels 
$\gamma h$ (blue curves), $\gamma Z$ (red curves) and 
$\gamma Z^\prime$ (purple curves) along with the continuum $\gamma$ ray emission (black
curves). Also shown are 95\%C.L. exclusion limits inferred from Fermi data\cite{Murgia} for a $\gamma h$ line
with $M_h=115$ GeV and a NFW profile.}
\label{fig:sigmav-vs-M}
\end{center}
\end{figure}

The expressions for ${\cal V}_{\gamma Z}^2$ and ${\cal V}_{\gamma Z^\prime}^2$
are given in the appendix.
The respective cross sections are then obtained by substituting Eqs. (\ref{eq:Msq-ZA})
and (\ref{eq:Msq-ZpA}) into Eq. (\ref{eq:sigma-v}).  In Fig.~\ref{fig:sigmav-vs-M}, we
plot the cross sections for $\gamma h$, $\gamma Z$ and $\gamma Z^\prime$ as a function 
of the neutrino mass $M_{\nu}$ for $g_\nu = g_t = 3$ 
and for two values of $M_{Z^\prime}$. We compare them with the continuum obtained 
from all 2-body annihilations.  

\subsection{Gamma Ray Spectra}

\begin{figure}[htb!]
\begin{center}
\includegraphics[width=0.6\textwidth]{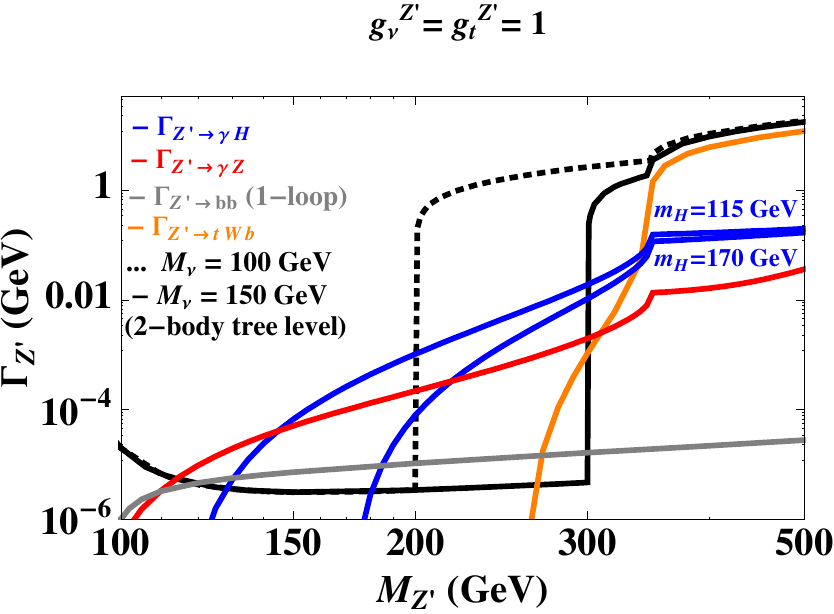} 
\caption[]{\small $Z^\prime$ partial widths as a function of its mass for
two values of the WIMP mass. 
The  solid and dotted black lines are for the sum of the partial widths into all 2-body tree level modes, which, apart from the $t\overline{t}$ channel, are induced by the kinetic mixing $\eta$. The gray line is for the one-loop annihilation into $b\overline{b}$, estimated using Eq.~\ref{eq:gzpbb}. One-loop annihilation into $\gamma h$ and $\gamma Z$ are shown in blue and red respectively. The orange line is for the three-body annihilation into $tWb$ .}
\label{fig:width}
\end{center}
\end{figure}

The photon spectrum originating from the process
$\nu \overline{\nu} \rightarrow \gamma X$ 
deviates from a monochromatic emission due to the finite decay width $\Gamma_X$ of the unstable particle $X$ with mass $M_X$
and is given by \cite{Bertone:2009cb}:
\bea
\label{eq:linespectra}
\frac{dN_{\gamma}^X}{dE}=\frac{4M_{\nu}M_X \Gamma_X}{f_1 f_2}
\eea
where
\begin{eqnarray}
f_1&=&\tan^{-1}\left(\frac{M_X}{M_{\nu}}\right)+\tan^{-1}\left(\frac{4M_\nu^2-M_X^2}{M_X\Gamma_X}\right)\nonumber \\
f_2&=&(4M_\nu^2-4M_\nu E_{\gamma}-M_X^2)^2+\Gamma_X^2M_X^2 \,. \nonumber
\end{eqnarray}
The Higgs width $\Gamma_h$ will match the SM prediction at tree level, but at loop level
could have tiny contributions from $h \rightarrow \gamma Z^\prime$ or
$h \rightarrow Z^\prime Z^\prime$, depending on $m_h$ and $M_{Z^\prime}$.
We neglect these insignificant corrections in the inclusive width.
The tree level $Z^\prime$ width is shown in Fig.~\ref{fig:width}, although
the final results are 
not sensitive to its precise value except very close to the $Z^\prime$ resonance.
\begin{figure}[htb!]
\begin{center}
\includegraphics[width=0.49\textwidth]{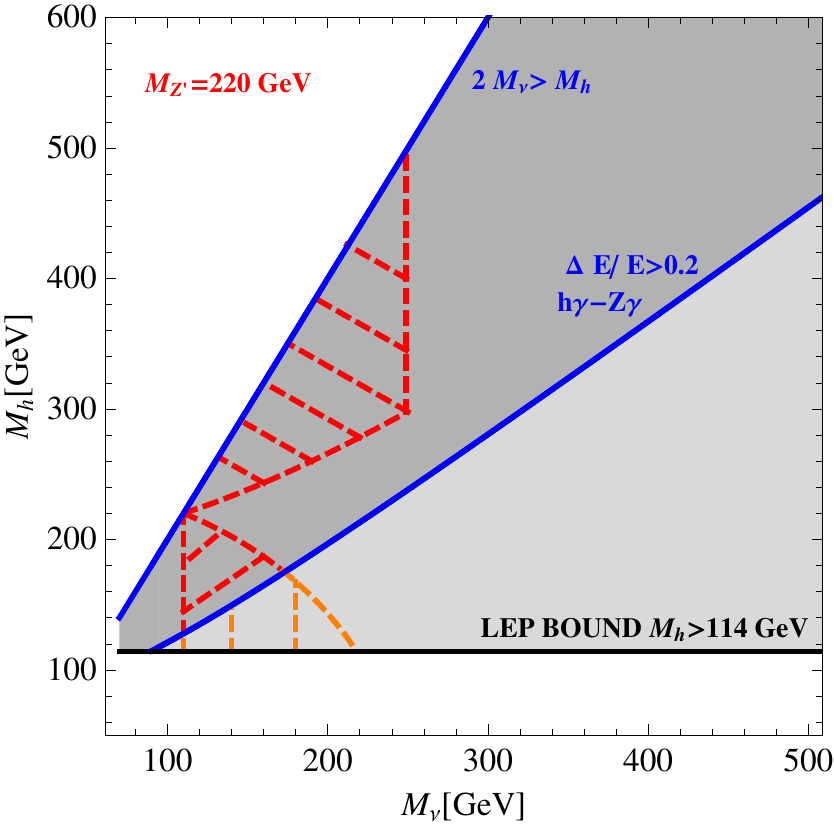} 
\includegraphics[width=0.49\textwidth]{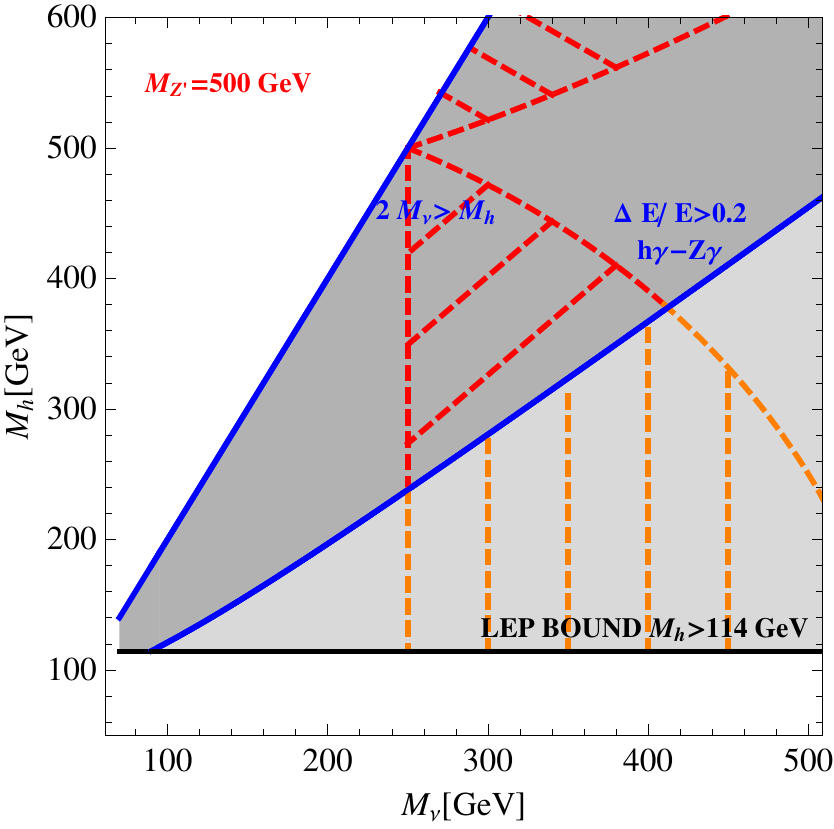} 
\caption[]{\small 
Regions of the $M_\nu$-$m_h$ parameter space 
(for $M_{Z^\prime} = 220$~GeV/500 GeV in the left/right panels)
for which the
$\gamma Z$ and $\gamma h$ lines
can be distinguished by an experiment with $10\%$ energy resolution (dark grey); in
the light grey region they are merged.  The red dashed area further shows the
regions where the
$\gamma Z$, $\gamma h$, and $\gamma Z^{\prime}$ lines are distinguishable. In the dashed
orange region $\gamma Z$ and $\gamma h$ lines are merged but distinguishable from the
$\gamma Z^{\prime}$ line.}
\label{fig:Mnu-vs-Mh}
\end{center}
\end{figure}

The detection of gamma ray lines {\it{per se}} represents
smoking-gun evidence for dark matter annihilation, but 
it does not tell us which processes are responsible for the
observed lines.
However, additional indirect dark matter searches,
direct detection experiments, and LHC observations
can complement
the information from gamma-ray telescopes.
For example, the energies of gamma-lines probe
the masses of the particles in the associated annihilation 
process, c.f. Eq.~(\ref{eq:energy}),
and this could be combined with independent
measurements of particle masses at colliders.
This cross-check could prove extremely useful to
identify a given long-lived particle produced at colliders as
a significant fraction of the dark matter present in the galaxy.

The detection and identification of
the $Z\gamma$ and $h\gamma$ lines could also allow
one to determine the Higgs mass.
Fig.~\ref{fig:Mnu-vs-Mh} shows the 
region in the $M_{\nu}$-$m_h$ plane where these two lines are potentially
separately observable.
The $h\gamma$ line can be distinguished from
the $Z\gamma$ line if the energy separation is at least
twice the energy resolution of the experiment, which for the
Fermi LAT is $\delta E\sim 10\%$
for the energies of interest \cite{Atwood:2009ez}.
The maximum Higgs mass which can be
probed in $\nu \overline{\nu} \rightarrow h\gamma$ annihilation 
is $2 M_\nu$.
For $2M_\nu >M_{Z^{\prime}}$, the $Z^{\prime}\gamma$ line
is also present.
In Fig.~\ref{fig:Mnu-vs-Mh} we show, for the representative
cases of $M_{Z^{\prime}}=220$ GeV and 500 GeV, the combination
of Higgs and $\nu$ masses for which all three of the lines
are distinguishable by an experiment with $\sim 10\%$ energy resolution.

In most DM models producing line signals, there will typically be one line from annihilation into $\gamma \gamma$ and/or $\gamma Z$ (these two lines can be resolved only if the wimp mass is in the $\sim 50 - 100$ GeV mass range). Measuring the energy of this line will provide useful information on the DM mass. In addition, if another less energetic line is detected, this will allow to estimate the mass of the new heavy particle $X$ that DM annihilates into. Since we are considering gamma ray energies between a few GeV to a few hundreds of GeV, this means that the DM and $X$ particles will be kinematically accessible at the LHC (if heavier than a TeV the corresponding gamma ray signal will be suppressed). Therefore any line signal observed with FERMI, MAGIC or HESS should be accompanied by a signal at the LHC.

\subsection{Dark Matter Distribution}
\label{sec:distribution}

The differential photon flux produced by dark matter annihilations
and collected from a region of angular size $\Delta \Omega$ is computed as:
\bea
\label{eq:gammaflux}
\Phi_{\gamma}(E_{\gamma}) & = & \frac{1}{4 \pi}\frac{r_{\odot}\rho_{\odot}^2}{4 M_{\nu}^2} \frac{dN_{\gamma}}{dE_{\gamma}} \bar{J}\Delta \Omega
\eea
with
\bea
\label{eq:cx-gammaflux}
\frac{dN_{\gamma}}{dE}&=&\sum_{f}\langle \sigma v\rangle_f \frac{dN_{\gamma}^f}{dE} \hspace{1cm}
\bar{J}=\frac{1}{\Delta \Omega}\int_{\Delta \Omega} J(\psi) \nonumber \\
J(\psi)&=&\int_{los} \frac{ds}{r_{\odot}} 
\left( \frac{\rho(r(s,\psi))}{\rho_{\odot}}\right)^2 .
\eea
$dN_{\gamma}/dE$ includes all possible annihilation final states $f$,
with $\langle \sigma v\rangle_f$ and $dN_{\gamma}^f/dE$
referring to the corresponding cross sections and photon spectra per annihilation.
The factor $1/4$ in Eq.~(\ref{eq:gammaflux}) is appropriate for
a Dirac fermion WIMP with predominantly particle-anti-particle annihilation modes.
The dimensionless quantity $J(\psi)$ 
corresponds to the integration of the photon signal along a line
of sight making an angle $\psi$ with the direction of the galactic center.
The total observed flux is then obtained integrating the emission over the 
the observed region of angular size $\Delta \Omega$.
The normalization factors $\rho_{\odot}=0.3$ $\mbox{GeV }\mbox{cm}^{-3}$ 
and $r_{\odot}=8.5$ kpc correspond 
respectively to the dark matter density at the solar position and to 
the distance of the Sun from the galactic center.
We model the dark matter density distribution in our galaxy, $\rho(x),$ 
as a Navarro-Frenk-White (NFW) profile ~\cite{Navarro:1995iw},
which is a good fit to current $N$-body simulations:
\begin{equation}\label{eq:NFW}
\rho_{\rm{NFW}}(r) = \frac{\rho_s}{\frac{r}{r_s}\left( 1 +
\frac{r}{r_s}\right)^2} \,. \end{equation}
Some recents simulations, however, prefer the so-called ``Einasto''\cite{Navarro:2008kc} 
profile, which is slightly more shallow at small radii and it does not
converge to a definite power-law:
\begin{equation} \rho_{\rm
Einasto}(r) = \rho_s\cdot \exp\left[- \frac{2}{\alpha} \left(
\left(\frac{r}{r_s}\right)^\alpha -1 \right) \right], \qquad
\alpha=0.17 \,.
\end{equation}

\begin{center}
\begin{table}
\centering
\begin{tabular}{|c|ccc|}
\hline
MW halo model &$r_s$ in kpc&$\rho_s$ in GeV/cm$^3$&$\bar J\, \left(10^{-5}\right)$ \\
\hline
NFW~\cite{Navarro:1995iw} & 20 & 0.26 & $15\cdot 10^3$ \\
Einasto~\cite{Navarro:2008kc} & 20 & 0.06 & $7.6 \cdot 10^3$\\
Adiabatic\cite{Bertone:2005hw} &  & & $(1.2-14.0) \cdot 10^7 $\\
\hline
\end{tabular}
\caption{Parameters of the dark matter density profiles for the Milky Way
discussed in the text and corresponding value of $\bar{J}$ for
$\Delta\Omega=10^{-5}$.} \label{tabprofiles}
\end{table}
\end{center}

The presence of baryons, not accounted for in the simulations previously quoted,
may significantly change the picture, particularly in the inner region
of the galaxy where the gravitational influence of the super massive black hole
is expected to have a large feedback on the surrounding dark matter distribution.
The evolution of the dark matter density profile, accounting for dark matter-star interactions,
capture in the central black-hole and the presence of dark matter annihilations, 
has been simulated in the so-called ``adiabatic compression'' scenario in Ref.\cite{Bertone:2005hw}.  
The final density distribution is significantly increased at 
small radii with respect to the initial NFW profile.

In Table \ref{tabprofiles} we show the $\bar{J}$ factor for the dark matter distributions
discussed here for an observation of the galactic central region with an 
angular acceptance $\Delta \Omega=10^{-5},$ corresponding to the angular
resolution of Fermi-LAT and current Air Cherenkov Telescopes (ACTs).
The adiabatically contracted profiles of Ref.\cite{Bertone:2005hw}
depend on the wimp annihilation cross section and in Table \ref{tabprofiles} 
we list the range of $\bar{J}$ factors that have been used 
in Fig.\ref{fig:photonfluxes1}.
The large uncertainties in the dark matter distribution in the region considered
turn into large uncertainties on the predicted photon fluxes.
On the other hand the $\rho^2$ dependence of the signal suggests the galactic central
region as the best target to maximize the signal.
For the rest of the paper we adopt NFW as a dark matter profile benchmark 
and for illustration we show some of the predictions for the adiabatically-contracted
profile.

%results can easily be rescaled for other profiles, as in Tab. \ref{tabprofiles}.

\subsection{Predicted Photon Fluxes and Comparison with Experiments}

The expected photon signal, $\Phi_{\gamma}^S$ is obtained by 
convolving the photon flux in Eq.~\ref{eq:gammaflux} with the energy response of
the instrument $G(E_0,E)$:
\bea
\label{eq:energyresolution}
\Phi_{\gamma}^S(E) & = & \int dE_0 \Phi_{\gamma}(E_0) G(E_0,E)
\eea
where we assume a Gaussian kernel
\bea
\label{eq:kernel}
G(E_0,E) & = & \frac{1}{\sqrt{2\pi}E\sigma} \exp \left(-\frac{(E_0-E)^2}{2\sigma^2E^2}\right)
\eea
with $\sigma$ depending on the detector energy resolution $\xi$ as $\sigma=\xi/2.3.$

\begin{figure}[h!]
\begin{center}
\includegraphics[width=0.357\textwidth]{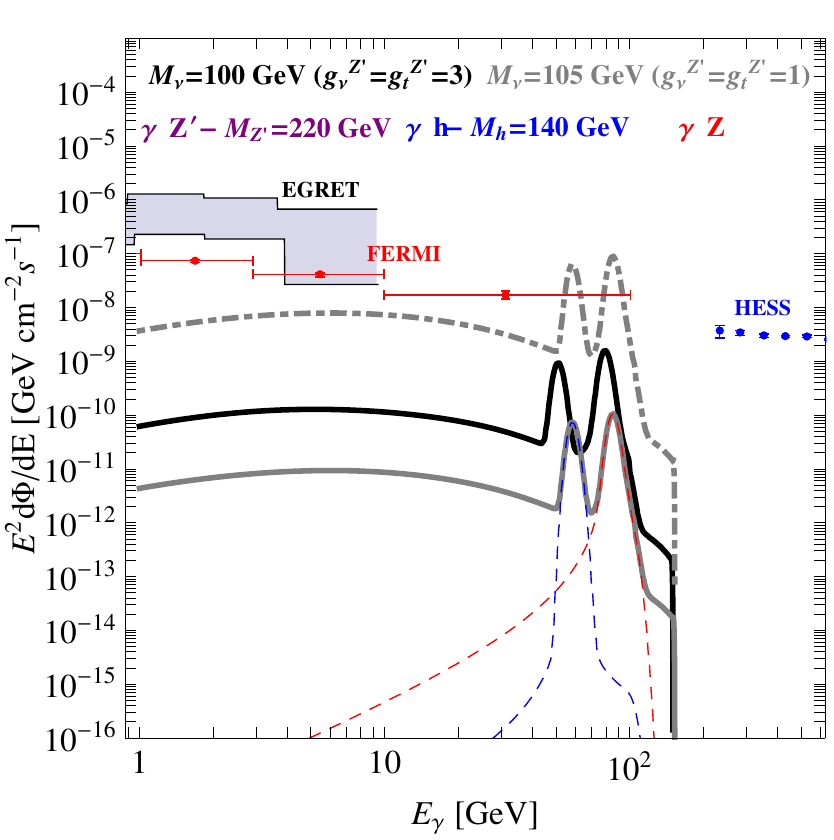}
\includegraphics[width=0.357\textwidth]{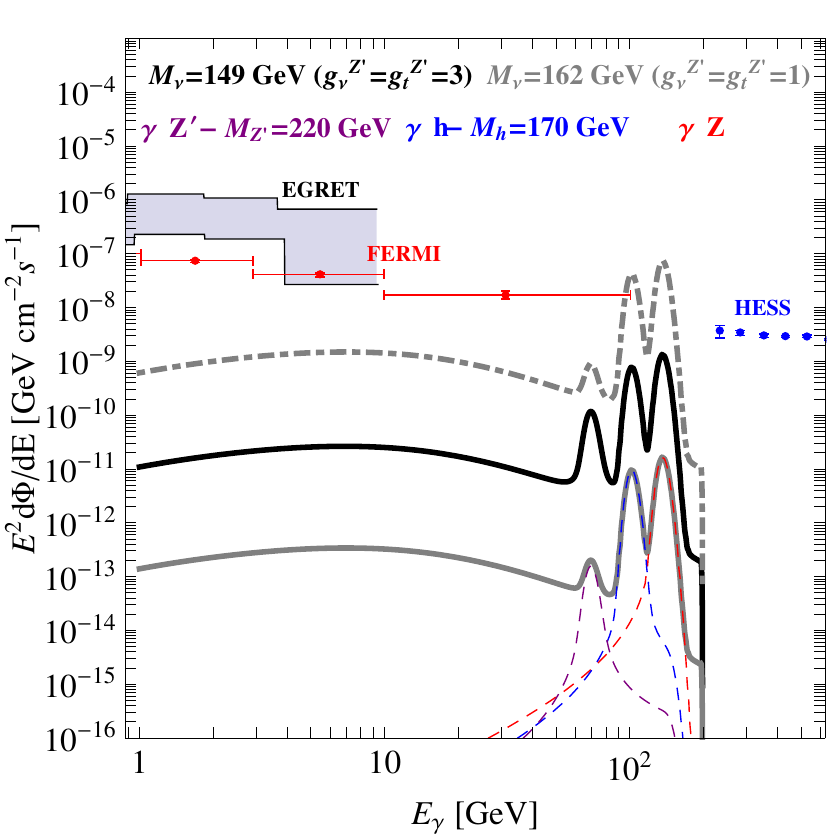}
\includegraphics[width=0.357\textwidth]{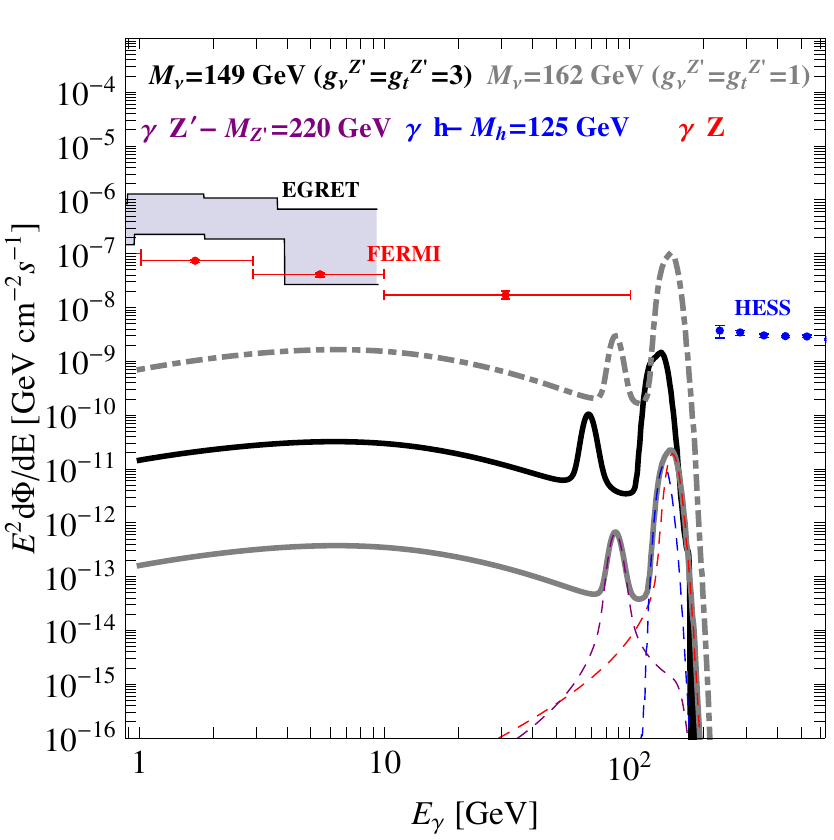}
\includegraphics[width=0.357\textwidth]{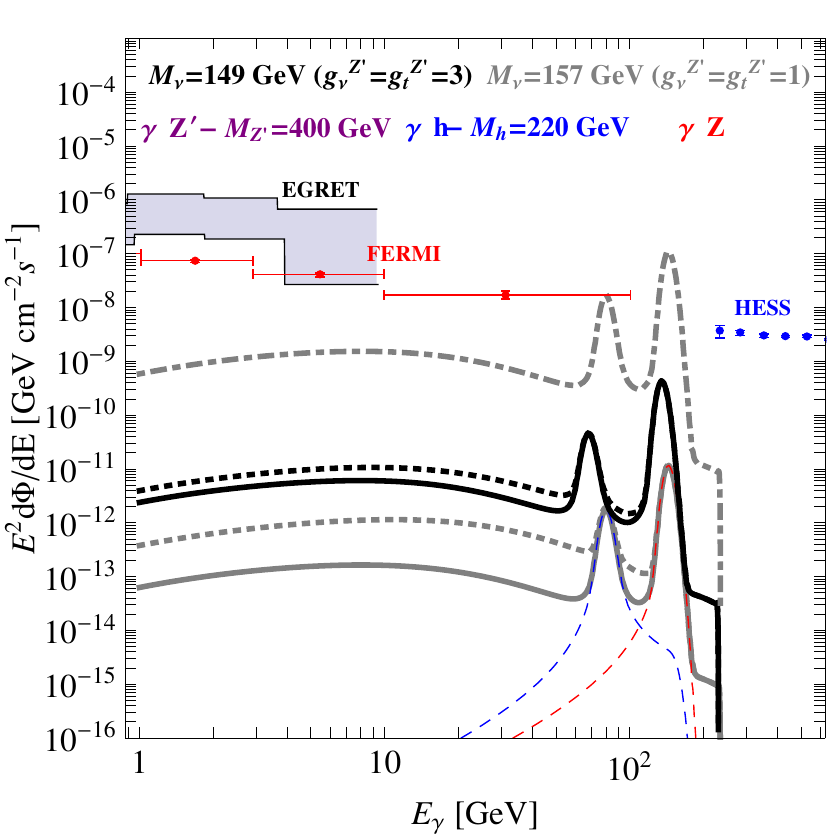}
\includegraphics[width=0.357\textwidth]{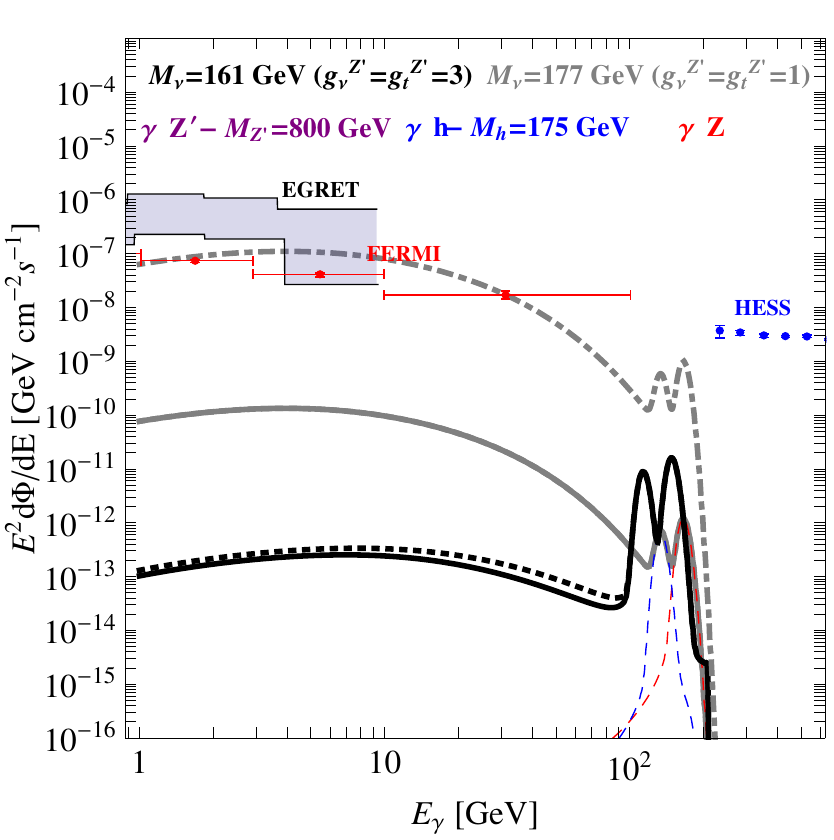}
\includegraphics[width=0.357\textwidth]{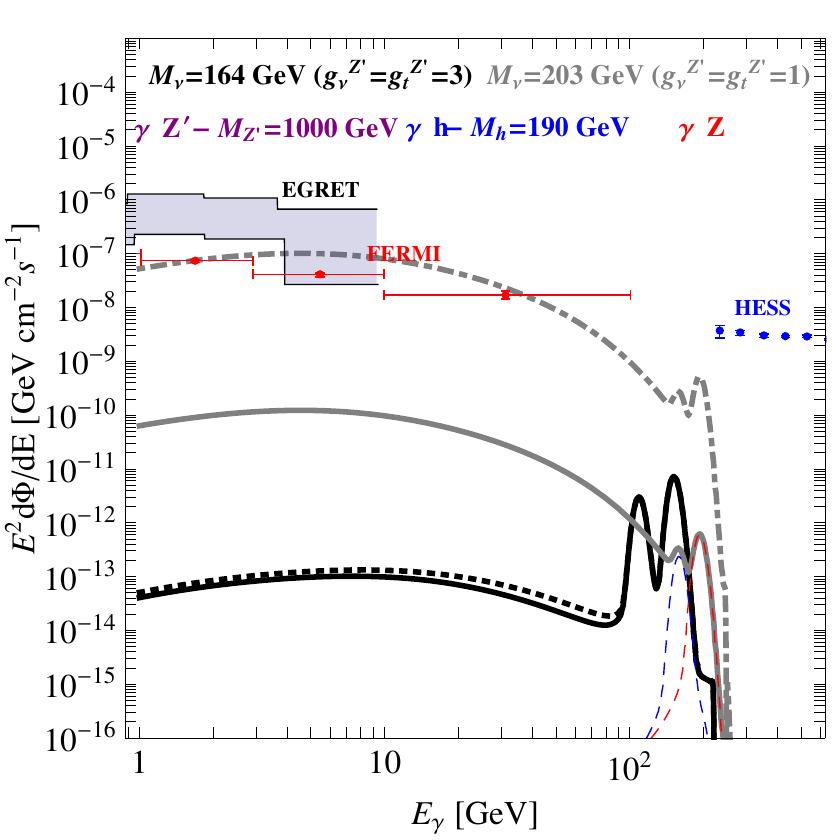}
\caption{\small Spectra obtained for different choices of mass parameters
and coupling $g^{Z^{\prime}}_{\nu,t}=1$ (black), $g^{Z^{\prime}}_{\nu,t}=3$ (gray), 
$\eta=10^{-3}$ (solid), $\eta=10^{-2}$ (dotted) that lead to the correct relic density and 
satisfy direct detection constraints. 
Upper plots are for $\eta=10^{-3}$ only since for these choices
of couplings and $M_{Z^{\prime}}$ mass $\eta=10^{-2}$ is excluded by direct detection constraints.
$\Delta \Omega=10^{-5}$, and a NFW dark matter 
profile is assumed. Dot-dashed lines are for the adiabatically-contracted
profile in Table \ref{tabprofiles}, $\Delta \Omega=10^{-5}$,  $g^{Z^{\prime}}_{\nu,t}=1$
and $\eta=10^{-3}.$ EGRET data are from \cite{MayerHasselwander:1998hg,Hartman:1999fc},
HESS from \cite{Aharonian:2009zk} and Fermi from \cite{Cohen-Tanugi}.}
\label{fig:photonfluxes1}
\end{center}
\end{figure}

In Fig.~\ref{fig:photonfluxes1},
we show the predicted photon fluxes at the galactic center for
different choices of particle physics parameters which give the correct
thermal relic abundance and satisfy the constraints from direct detection. 
For comparison we plot
the HESS observations of the same angular region \cite{Aharonian:2009zk} and
the EGRET data on the unidentified source 3EG J1746-2851\cite{MayerHasselwander:1998hg,Hartman:1999fc}, corresponding 
instead to  $\Delta \Omega=10^{-3},$ appropriate for a detector 
with an angular resolution of  $\sim 1^{\circ}$. 
Fermi satellite preliminary results fill the region in energy between 
HESS and EGRET, providing the
most powerful probe of WIMP annihilation into gamma rays to date.

The Fermi collaboration has recently presented 
preliminary $95\%$ C.L. upper limits on the line cross section 
from a line search based 
on a shape analysis \cite{Cohen-Tanugi} of 11 months of observation 
of a region excluding the galactic plane but including the galactic central region. 
%(all sky - $|b|<10^{\circ}$ + $\psi<30^{\circ}$).  
We translate the preliminary bounds 
on $\gamma \gamma$ and $\gamma Z$ to bounds on $\gamma X$ for 
$M_X = 115$~GeV.  Comparing even the current bounds 
with the $\gamma h$ cross sections in Fig.~\ref{fig:sigmav-vs-M}, we see that Fermi is 
{\em already} ruling out a small region of parameter space where the annihilation 
is close to on-resonance.  Thus, even though its line search so far has had a 
null result, longer exposure and optimized windows of observation still have major
potential to observe a signal. 
Emission from dark matter is expected to be maximized at the galactic center, 
but that may not be the optimal region to detect this signal. 
For example, the intense point-like gamma-ray source at the galactic center 
detected by the ACTs at energies 
above $200$ GeV and confirmed by Fermi at lower energies, 
constitutes a serious background for dark matter searches. 
The morphology of the background and signal emissions suggests that a better 
strategy could be to consider a larger angular region, with the size 
depending on the choice of dark matter profile (e.g.\cite{Serpico:2008ga}) and subtract all 
the astrophysical point sources there detected. 
This kind of analysis is ongoing within the Fermi collaboration \cite{Murgia}. 

In Fig.\ref{fig:photonfluxes1} we are not exhausting all the possible predictions of our model.
The spectra  shown are instead meant to be illustrative cases which capture different 
phenomenological aspects of the theory and present some of the possible outcomes rather than  characterize all of them. 
For instance, we see how, depending on the relative masses of the WIMP and the Higgs, there can be one, two (either $\gamma$-$h$ and $\gamma$-$Z$ or if these two are merged, $\gamma$-$Z^\prime$ and $\gamma$-$h/Z$) or three lines and 
for  WIMP masses below the top mass, as it  happens in the large coupling regime, there are spectra with prominent lines compared to the continuum. 
Taking also into account the uncertainties in the dark matter distribution, which we illustrate  by considering two different profiles, 
it is clear that the line signals we are presenting may be at the reach of future Fermi observations. 
Besides Fermi, the next generation of low-threshold, air cherenkov telescopes, such as CTA \cite{CTA} may potentially have a  dramatic impact on dark matter searches. 
In particular,  they may probe the line signals shown in Fig.\ref{fig:photonfluxes1} due to their improved 
energy threshold, which is expected to be lowered 
to $\sim 30$ GeV\footnote{Considering a future CTA observation of a $2^{\circ}$ region around 
the galactic center and based on a signal-to-noise analysis, 
we have sketched the 3$\sigma$ evidence prospects (assuming a NFW profile) 
for a generic line $\gamma X$. We have assumed a CTA effective area of $\sim 1$ km$^2,$ 
an energy resolution of 15\% and 200 hours of observations. 
The prospects strongly depend on the hadronic rejection efficiency, which 
we take energy independent and we vary from a conservative value of $ \epsilon= $50\%
to a more optimistic $ \epsilon= $1\%.
For example, focusing on a $\gamma X$ line with $M_X=115$ GeV and for $M_{\nu}=100$ GeV 
the cross-section at the reach of CTA are $\sim 2$ $10^{-28}$  cm$^3$ s$^{-1}$ 
$(\epsilon=1\%)$ and $\sim  10^{-27}$  cm$^3$ s$^{-1} (\epsilon=50\%)$. }.

\section{Other Signals and Constraints}
\label{sec:othersignals}

\subsection{Elastic Scattering and Direct Detection}
\label{sec:direct}

\begin{figure}[tbh]
\begin{center}
\includegraphics[width=0.4\textwidth]{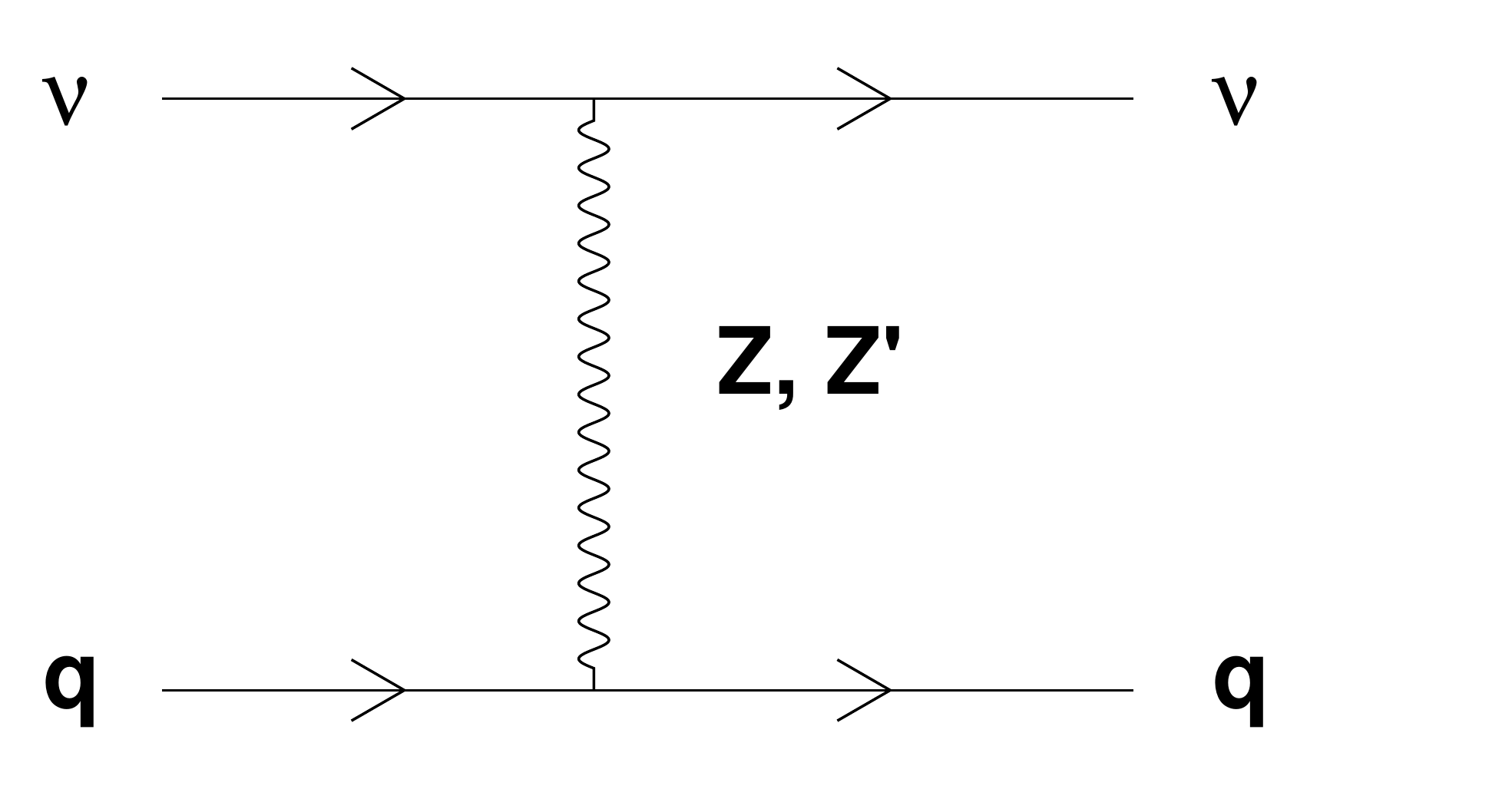} 
\caption[]{\small Representative diagram indicating how $\nu$ may scatter elasticly 
with nuclei.}
\label{fig:elastic}
\end{center}
\end{figure}

The fact that there has been no unequivocal observation to date of dark matter
scattering with heavy nuclei places a constraint on any theory of dark matter and, in particular,
results in a bound on the $Z$-$Z^\prime$ mixing parameter $\chi$ of
Eq.~(\ref{eq:langrangian}).  As shown in Fig.~\ref{fig:elastic}, $\nu$ scatters with
nuclei largely by exchanging a $Z$ or $Z^\prime$ boson which couples to valence quarks.
This is a direct consequence of the fact that $\nu$ is a Dirac fermion, which
(unlike a Majorana one) generally has vector interactions which 
remain large in the non-relativistic limit.  As a result, the non-observation of a signal
at direct detection experiments puts strong constraints on the couplings involved in the
reaction. The $\nu$-nucleon elastic scattering cross section is given by:
\bea
&\sigma_{\stackrel{neutron}{proton}}&=\frac{g^2 \ m_{n/p}^2 \  {g_{\nu}^{Z^{\prime}}}^2 }{64\pi  c^2_W}
\left[  \frac{t_{\alpha}c_{\alpha}}{M_Z^2} \left(\stackrel{1-s_W t_{\alpha} \eta}{1-4s^2_W+3\eta t_{\alpha}s_W}\right)
-\frac{A_{\alpha}}{M^2_{Z^{\prime}}} \left(\stackrel{1}{1-4s^2_W  B_{\alpha}}\right)
\right]^2 \, ,
\label{eq:spin_dependent}
\eea
where,
\begin{equation}
A_{\alpha}= c_{\alpha} (t_{\alpha}+\eta s_W) \ \ ~{\rm and}~ \ \ 
B_{\alpha}=(t_{\alpha}+\eta/ s_W)/(t_{\alpha}+\eta  s_W).
\end{equation}
Elastic scattering of $\nu$ with target nuclei is entirely on protons because the
scattering on neutrons vanishes due to a cancellation between the 
$Z$ and $Z^{\prime}$ contributions.  The proton scattering is
almost entirely from the $Z^{\prime}$ exchange. 

\begin{figure}[tbh]
\begin{center}
\includegraphics[width=0.7\textwidth]{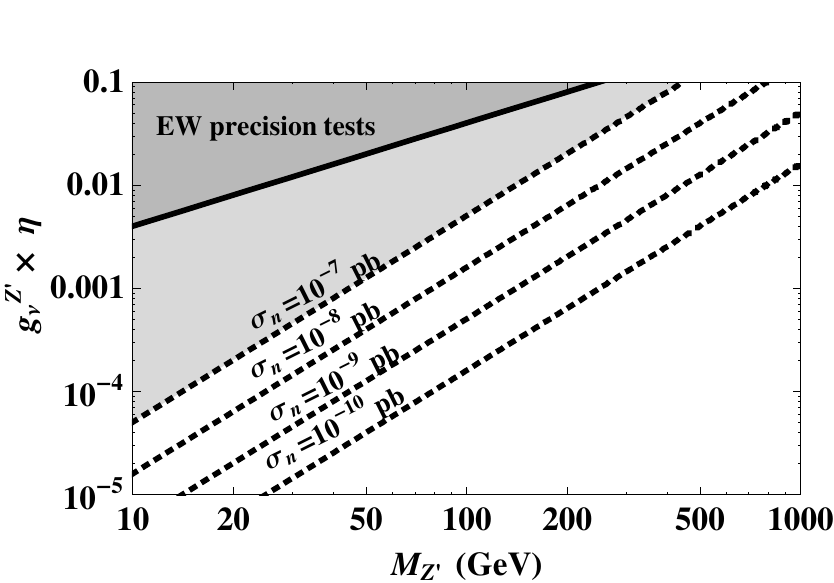} 
\caption[]{\small Contours for the $\nu$-nucleon elastic scattering cross section. The dark grey region is excluded by EW precision data. The lighter grey region indicates the approximate region ruled out by CDMS (the precise limit depends on the WIMP mass).}
\label{fig:CDMS}
\end{center}
\end{figure}

 In Fig.~\ref{fig:CDMS} we show the upper limit on the product of
$g_{\nu}^{Z^\prime} \times \eta$ based on the null result of the latest CDMS
search for elastic WIMP-Germanium scattering \cite{Ahmed:2008eu}.  
The dependence on the $\nu$ mass is relatively mild provided it is greater than
the mass of Germanium, about 70~GeV.
Provided the 
$Z^\prime$ mass is larger than a few tens of GeV, the constraints are consistent with 
order one coupling between $\nu$ and the $Z^\prime$, and $\eta$ consistent with
the expectation that it is induced at the one loop level. Note that these constraints are stronger than those from the EW precision measurements
\bea
 \left(\frac{\eta}{0.1} \right)^2
 \left(\frac{250 \mbox{ GeV}}{M_{Z^\prime}}\right)^2  & \lesssim & 1,
\eea
derived in \cite{Kumar:2006gm}.

\subsection{Relic Density}
\label{sec:relic}

As a Dirac fermion, $\nu$ can carry a conserved $U(1)_{\nu}$ global charge, raising the
possibility that its relic density may be understood as a primordial asymmetry
between WIMP particles and antiparticles\footnote{Of course, annihilation into
gamma rays today requires a small breaking of $U(1)_\nu$ such that
particles and antiparticles re-equilibrate after freeze-out.}.  However, even in the absence of a 
$\nu - \bar{\nu}$ asymmetry, $\nu$ has roughly the right properties for its thermal relic abundance
to be appropriate to match cosmological measurements.  Thus, while we present the
relic density as a very interesting way to understand the parameter space, it is not
a firm bound in the sense that the density of $\nu$ today
may be due to a primordial asymmetry or a nonstandard cosmology.  In fact,
we do find regions with the correct thermal relic abundance which can lead to beautiful 
gamma ray line signatures at Fermi.

\begin{figure}[t]
\begin{center}
\includegraphics[width=0.485\textwidth]{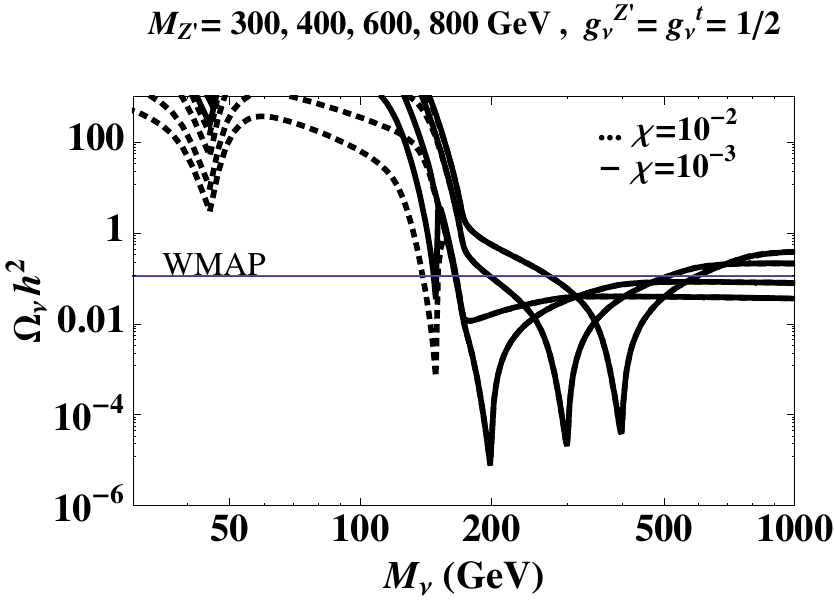} 
\includegraphics[width=0.485\textwidth]{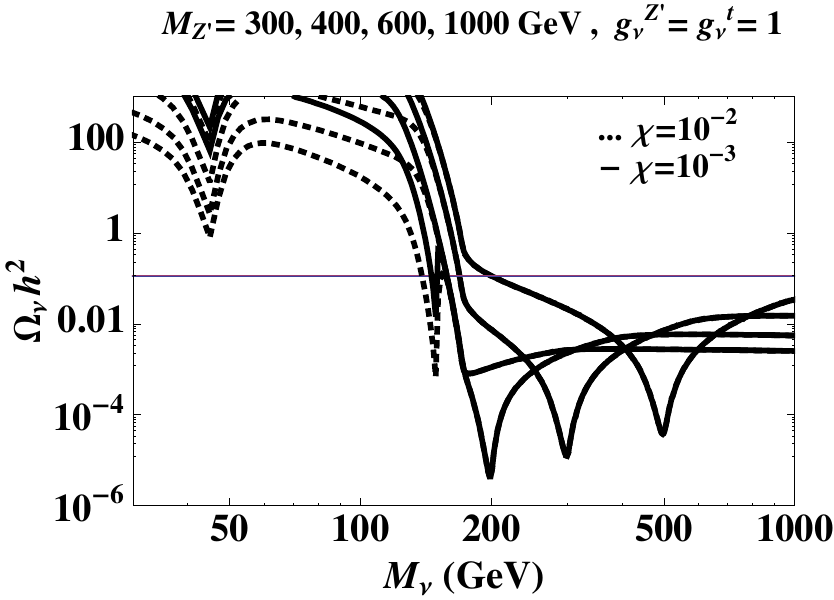} 
\includegraphics[width=0.485\textwidth]{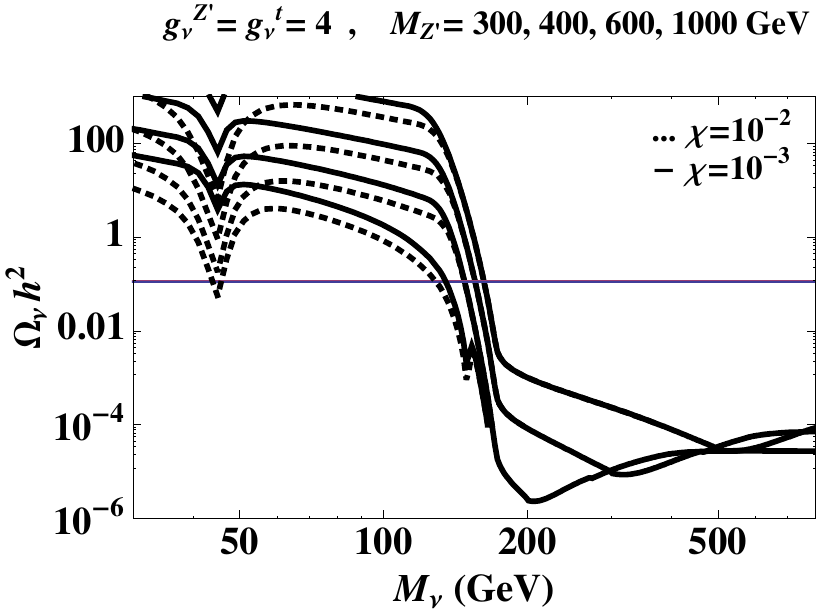} 
\includegraphics[width=0.485\textwidth]{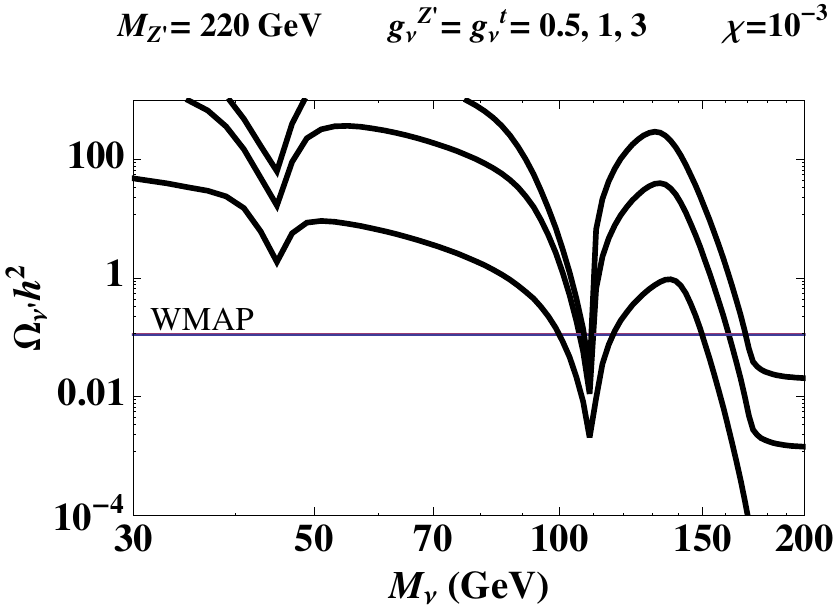} 
\caption[]{\small Relic density as a function of the $\nu$ mass for the indicated 
parameters. As the $Z^{\prime}$ coupling to top and $\nu$ increases, the prediction for 
the $\nu$ mass gets narrower assuming $\nu$ is the sole contributor to the relic density: 
For $g_{\nu}^{Z^{\prime}}, g_t^{Z^{\prime}} \gtrsim 1 $ this implies $  M\sim 150$ GeV, 
whatever the value of $ M_{Z^{\prime}} \gtrsim 400 $ GeV. For lighter $ M_{Z^{\prime}}$, 
several $M_{\nu}$ can lead to the correct relic density.
The calculation includes all 2-body tree level annihilations as well as the 
loop-induced annihilations into $b\overline{b}$.}
\label{fig:relic}
\end{center}
\end{figure}

The abundance of a thermal relic $\nu$ is controlled by
its annihilation cross section into SM particles, 
mediated by the same diagrams which control annihilation into gamma rays today shown in
Fig.~\ref{fig:anni_diagrams}.  
To circumvent the constraints from elastic scattering in Fig.~\ref{fig:CDMS}, 
we restrict our discussion to $M_{Z^{\prime}}\gtrsim$ 200 GeV, for which 
the results can be broadly classified:
\begin{itemize}

\item For $M_{\nu} \lesssim M_Z/2$ , the annihilation proceeds largely
through the $\nu$ coupling to $Z$, into light SM fermions and is 
controlled by the size of the kinetic mixing. 
The relic density is typically in conflict with direct detection constraints unless 
the annihilation is very close to the resonance, and the small WIMP mass
kinematically forbids all of the line signals.

\item For $M_Z/2 \lesssim M_{\nu} \lesssim m_t$,
annihilation is  via loop-level processes into 
$\gamma Z$, $\gamma h$ and $b \overline{b}$ (for $M_\nu \sim m_t$ there is
also some annihilation into off-shell $t \bar{t}$).  
Far below the
$Z^\prime$ resonance for modest $\nu$ coupling to the $Z^\prime$,
the rate is typically too small, leading to over-abundance of $\nu$ in the early Universe.
This is remedied by having $M_\nu$ slightly above or below $M_{Z^\prime} / 2$
or large $Z^\prime$ couplings, for which the relic density prefers
$M_\nu$ slightly smaller than $m_t$.

\item For $M_{\nu}\gtrsim m_t$, the non-relativistic annihilation proceeds 
largely into a $t \bar{t}$ final state. 
For moderate couplings, $g_{\nu}^{Z^\prime}\sim g_{t}^{Z^\prime} \sim 1/2$, there is a continuum 
of possibilities which are highly correlated with the position of the $Z^\prime$ resonance.
As the coupling is dialed stronger this window becomes narrower and occurs for $\nu$ masses
close but smaller than the top mass.
Taken together with the previous case, this is a 
robust prediction of the model in the strong coupling regime. 
Whatever the value of $M_{Z^{\prime}}\gtrsim 350 $ GeV, the 
DM mass is around 150 GeV, far away from the $Z^\prime$ resonance. 

\item Finally, if $M_{\nu} \gsim M_{Z^\prime}$,  the $t$-channel 
process $\nu \overline{\nu} \rightarrow Z^\prime Z^\prime$ opens up.  However, this does 
not play an important role since we are considering $M_{Z^\prime}\gtrsim 200$ GeV and 
in this case the annihilation cross section into top quark pairs continues to dominate.  

\end{itemize}
The predicted thermal relic density for several representative parameter sets are shown
in Fig.~\ref{fig:relic}. The calculation was done with MicrOMEGAs \cite{Belanger:2006is}.

\subsection{Anti-matter Signals and radio constraints}

\begin{figure}[htb!]
\begin{center}
\includegraphics[width=0.49\textwidth]{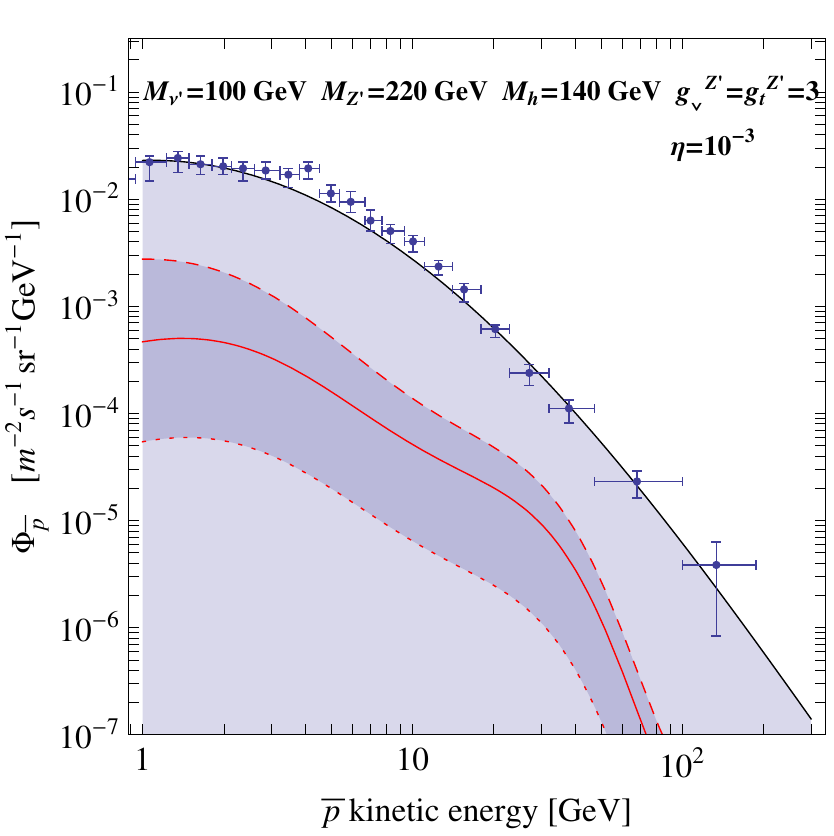}
\includegraphics[width=0.49\textwidth]{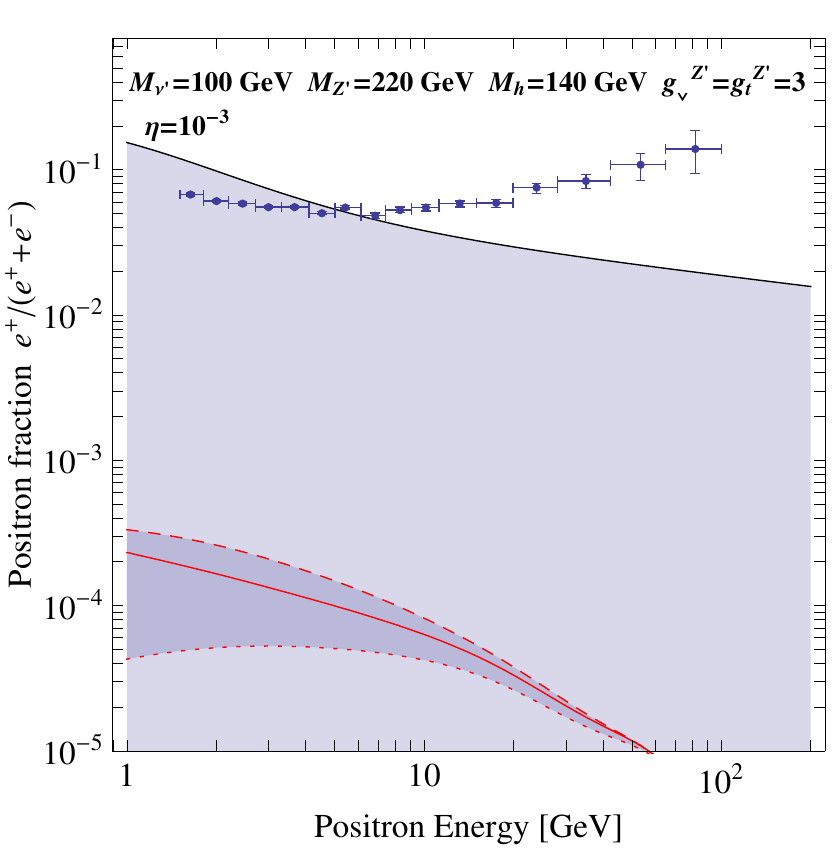}
\caption{\small Left: antiproton flux measured by PAMELA \cite{AdrianiCosmo09}.
Red lines show the contribution from $\nu$ annihilations for different sets of propagation parameters.
In particular, dashed, solid and dotted red curves are respectively for the MAX, MED and MIN
propagation models proposed in \cite{Donato:2003xg}.
Right: positron fraction measured by PAMELA \cite{Adriani:2008zr}. 
Dashed, solid and dotted red curves assume respectively the MAX, MED and MIN
propagation models in \cite{Delahaye:2007fr}.
In both panels, the black lines correspond to the expected astrophysical backgrounds.
Antiproton fluxes, computed for a NFW profile, and background estimates are obtained as
in \cite{Cirelli:2008id}.}
\label{fig:antiprotons}
\end{center}
\end{figure}

Other than photons, $\nu$ annihilations produce
$e^+$ and $\bar{p}$ fluxes which can leave their imprint in
the energy spectra of galactic cosmic-rays.
PAMELA $\bar{p}$ data can in particular set stringent constraints
on such contributions, as it has been noticed in Refs.~\cite{Cirelli:2008pk,Donato:2008jk, Bringmann:2009ca}.
We have computed $\bar{p}$ fluxes induced by tree level $\nu \overline{\nu} $
annihilations into SM particles as well as $\bar{p}$ produced by $h$ and $Z$ decays
associated with the $\gamma h$ and $\gamma Z$ line processes.
For each model in Fig.~\ref{fig:photonfluxes1} the antiproton signal is
well below the experimental data.
Generally, since relic density calculations usually predict $\nu$ masses of the order
$\mathcal{O} (100)$ GeV, we don't expect any features in the antiproton spectrum at energies
above present PAMELA measurements.
The same considerations hold for cosmic-ray positrons: the predictions fall
in the energy range explored by PAMELA and we find a very small $\nu$ contribution 
to the measured positron fraction, i.e. $e^+/(e^++e^-).$  While $\nu$ may not describe the PAMELA excess, several astrophysical processes such as pulsars~\cite{pulsars} or supernovae remnants~\cite{Blasi:2009hv} offer an explanation.
For illustration, in Fig.~\ref{fig:antiprotons} we show the predictions for $e^+$ and $\bar{p}$
signals for one model in Fig.~\ref{fig:photonfluxes1}.
Inverse Compton scattering of high energy electrons and positrons (produced by $\nu$ annihilations)
off  interstellar photons produce gamma-rays.
The constraints on such emission from EGRET data and preliminary Fermi results in Ref.\cite{Cirelli:2009vg,Meade:2009iu}
are easily satisfied
by the models in Fig.~\ref{fig:antiprotons}, because of the small $e^{\pm}$ fluxes
associated with $\nu$ annihilations.
Further bounds on the WIMP annihilation cross section are obtained comparing 
radio observations of the galactic center with the synchrotron emission induced
by the propagation of such high energy electrons and positrons
in the galactic magnetic field.
The constraints are powerful but strongly dependent on the choice of the DM 
density profile, as shown in Refs.~\cite{Bringmann:2009ca, Regis:2008ij, Bertone:2008xr, Bergstrom:2008ag}.
In particular, current N-body simulations cannot probe the DM
density distribution in the inner region of the galaxy $<{\mathcal O}$(100 pc).
For a simple extrapolation of the NFW density profile suggested by simulations at these small radii,
we checked that the combination of masses and
cross-sections considered in Fig.~\ref{fig:photonfluxes1}
are compatible with the radio bounds.
For the enhanced profile of Ref.\cite{Bertone:2005hw}, resulting from the DM adiabatic contraction
around the central black hole of our galaxy, the gamma-ray fluxes can be in tension with the data and the
associated synchrotron emissions are typically at odds with the radio bounds.
Note that these profiles are taken as representative cases of enhanced DM density
distributions in the inner regions of the galaxy.
However, the formation and evolution of a DM cusp at
the galactic center is quite uncertain and other DM distributions
falling between the two cases cases here considered, i.e. NFW and adiabatically contracted profiles,
are possible (see Ref.\cite{Bertone:2005hw, Prada:2004pi}).

\subsection{Signals at High Energy Colliders}
\label{sec:collider}

\begin{figure}[t!]
\begin{center}
\includegraphics[width=0.7\textwidth]{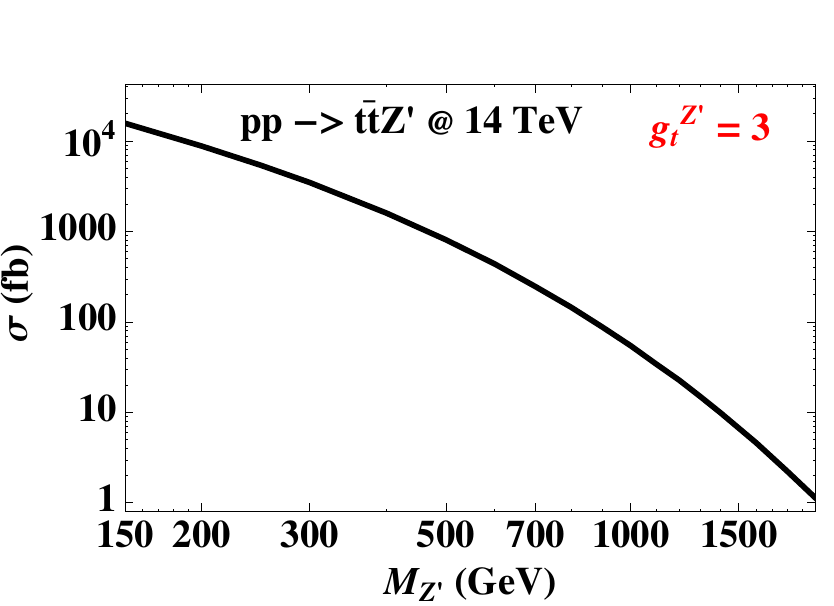}
\caption{\small Production cross section of $t\bar t$ in association with a $Z'$.  Above $M_{Z^\prime}> 2 m_t$, the branching fraction to top pairs can be large, resulting in a large $t\bar t t \bar t$ rate.}
\label{fig:LHC}
\end{center}
\end{figure}

Since the coupling of $Z^\prime$ to light SM fermions is suppressed by the 
small kinetic mixing factor, the best probe of the dark sector
is through the top portal.
In particular, the $Z^\prime$ can be produced by being radiated from top quarks, which
have a large QCD production cross section at hadron colliders.
In Fig.~\ref{fig:LHC}, we show the leading order cross section at LHC for 
$t\overline{t} Z^\prime$ production, as calculated by CalcHep \cite{Pukhov:2004ca}.
Depending on the masses and couplings, the $Z^\prime$ will predominantly decay
into $t \overline{t}$, $\nu \bar{\nu}$, or into light fermions.
Decays into top quarks lead to four-top events with a very large cross section compared
to the SM four top rate, which can be visible through a same-sign dilepton
signature \cite{Lillie:2007hd} (see also \cite{Contino:2008hi} for studies of a $ttWW$ final state). 
The right-handed nature of the $Z^\prime$ coupling to tops implies top polarization 
also provides an interesting observable.
When the $Z^\prime$ decays into WIMPs, a $t\overline{t} +$ missing energy final state results,
which presents a more challenging search at the LHC, but is definitely worth investigating. 
Work in these directions is in progress \cite{in progress}. 
When the $Z^\prime$ is light, it may have large decays through its loop-induced interactions
into $\gamma h$ or $b \bar{b}$.  The $\gamma h$ decay provides a novel
monoenergetic photon signature from 
$t\overline{t}Z^\prime \rightarrow t \overline{t} h \gamma$. 
The decay into $b\overline{b}$ offers the possibility to reconstruct 
$M_{Z^\prime}$ from the $b \bar{b}$ invariant mass of
$t\overline{t}Z^\prime \rightarrow t \overline{t} b \overline{b}$ events.
Finally, the SM Higgs phenomenology at LHC could be changed due to the mixing with the Higgs responsible for the breaking of $U(1)^{\prime}$. This was reviewed in \cite{Wells:2008xg}.  Besides, in our model, the heavy Higgs could be produced by gluon fusion through a top loop and lead to interesting signatures in $t\overline{t}$ production. 

%================================================
\section{Discussion and Outlook}
\label{sec:conclusions}

In this article, we have shown that if dark matter has a large coupling to the top quark and suppressed couplings to light standard model degrees of freedom, as 
is expected when the DM dynamics are intimately linked with electroweak symmetry breaking,  
the Higgs could be produced copiously in the galaxy in association with a photon 
from dark matter annihilations. 
The resulting photon spectrum possesses a  line whose energy reflects the mass of the 
Higgs and of the WIMP and does not arise if DM is a scalar or a Majorana fermion, thus
providing information about the WIMP spin and statistics.
 We have illustrated this phenomenon in a case in which the
 DM is composed of Dirac fermions 
 annihilating via an $s$-channel massive vector resonance ($Z^\prime$) 
 which couples strongly to the top quark, serving as a portal between dark matter and the
 Standard Model.
Our setup arises naturally in models of  ``partial compositeness" in which the top quark
acquires its large mass (after EWSB)  through large 
mixing with composite states in a new strong sector, as in 4d duals to RS Models.

For couplings of order ${\cal O}(1)$, the correct dark matter abundance is reproduced 
from the standard thermal relic density calculation if the dark matter mass is of order 
the top mass, typically in the 100 GeV -- 170 GeV mass range.  In the limit of strong coupling,
this feature is not strongly dependent on the mass of the $Z^\prime$. 
This is a perfect mass range for searches with the Fermi LAT for gamma rays from WIMP annihilations, which we find has very good prospects for a discovery in the near future.

In part, gamma ray lines are particularly important because (unlike a typical model of WIMP DM,
for which the photon continuum is usually much larger than the loop suppressed gamma ray 
lines), for $M_{\nu} < m_t$ (as is favored by the relic density in the strong coupling regime), 
the annihilation processes  at the origin of the continuum photon emission are 
themselves a one loop process into $b \bar{b}$, enhancing the relative prominence of
annihilation into $\gamma h$ and $\gamma Z$.  Even for
$M_{\nu}\gtrsim m_t$,  the continuum originating from annihilation into 
top quark pairs is rather soft, and the lines remain visible.

To illustrate Higgs production in space, we have worked with a simple representative
effective theory that captures properties of a general class of models with (some)
simplifying assumptions. 
It would be interesting to refine our predictions for some particularly well-motivated models 
of electroweak symmetry breaking which fall into this category, such as 
technicolor models \cite{Gudnason:2006yj} or composite Higgs models \cite{Contino:2003ve}.
In our effective theory construction, the WIMP effectively
couples to the Higgs through the top quark, which may turn out to
be ideal in terms of producing a
$\gamma h$ line signal.  
Another interesting possibility would be to consider gauge-higgs unification models where the dark fermions couple directly to $W$ gauge bosons \cite{Carena:2009yt,Haba:2009xu}. Annihilation into $\gamma h$ would then be mediated by a box diagram involving $W$'s with couplings dictated by the gauge symmetry of the model. 
One could also easily imagine DM coupling directly to the Higgs through higher dimensional operators.
In this work, we illustrated the $\gamma h$ signal from annihilating dark matter.  However, also well-motivated are models where dark matter decay is induced by higher dimensional operators suppressed by a high scale close to the GUT scale, which, as mentioned in the introduction, can lead as well to gamma-ray line signatures. In fact, after this paper was posted, Ref.~\cite{Arina:2009uq} appeared, which adequately completes our study by calculating the $\gamma h$ signal arising from decaying vector dark matter, leading naturally to an intense gamma-ray line.

As we explore the weak scale, we expect the dynamics of the electroweak 
breaking to be revealed.  It may be that its secrets already shine down from the sky, 
produced by dark matter annihilation or decay.

%================================================
\section*{Acknowledgments}

The authors are grateful for inspiration and conversation involving
Elliot Bloom, Marco Cirelli, Laura Covi, Abdelhak Djouadi, Michele Doro,  Jim Henson, 
Alejandro Ibarra, Emmanuel Moulin, Simona Murgia, 
Pasquale Serpico, Jing Shu, and James Wells. 
We also thank Genevi\`eve B\'elanger and Sasha Pukhov for helping with MicrOMEGAs.
This research is supported by the European Research Council Starting Grant Cosmo@LHC.
T. Tait is grateful to the SLAC theory group for their extraordinary
generosity during his many visits.
Research at Argonne National Laboratory is
supported in part by the Department of Energy under contract
DE-AC02-06CH11357.  Research at Northwestern University is
supported in part by the Department of Energy under contract
DE-FG02-91ER40684.  The work of M. Taoso is partly supported by the Spanish
grants FPA2008-00319 (MICINN) and PROMETEO/2009/091 (Generalitat
Valenciana) and the European Council (Contract Number UNILHC
PITN-GA-2009-237920).

%================================================
\appendix

\section{A Simple 4d UV Completion}
\label{sec:uvcompletion}

One particularly simple UV completion is to start with the Standard Model, treating all of
its fields (including $t_R$) as uncharged under $U(1)^\prime$.  We include SM singlets
$\nu_L$ and $\nu_R$ which are charged under $U(1)^\prime$ to play the role of the WIMP,
and in addition, a pair of
fermions $\psi_L$ and $\psi_R$, whose SM gauge quantum numbers are identical to
$t_R$, but with equal charges under $U(1)^\prime$.  In this framework, the additional
ingredients are vector-like, and thus the SM and mixed SM-$U(1)^\prime$ anomalies
are trivially absent.  Depending on the charges of $\nu_L$ and $\nu_R$, there may
still be $U(1)^{\prime 3}$ anomalies, but these can be simply cancelled by adding 
SM gauge singlet fermions to the dark sector.

To realize coupling of the $Z^\prime$ to the top quark, we consider the gauge invariant
masses and Yukawa couplings of the top-$\psi$ sector,
\bea
y H \bar{Q}_3 t_R  + \mu \bar{\psi}_L \psi_R + Y \Phi \bar{\psi}_L t_R
\eea
where $Q_3$ is the 3rd family quark doublet, $H$ is the SM Higgs doublet,
$\Phi$ is the Higgs field
responsible for breaking $U(1)^\prime$, $y$ and $Y$ are dimensionless couplings,
and $\mu$ is a gauge-invariant mass term for $\psi$.  In the regime 
$ y \langle H \rangle \ll \mu, Y \langle \Phi \rangle$ and $ \mu \sim Y \langle \Phi \rangle$, this system forms an ``inverted
top see-saw" in which the light eigenstate has mass $\sim  y \langle H \rangle$ and is composed mostly of $t_L$ and some fifty-fifty mixture of $\psi_R$ and $t_R$, which we identify as the
top quark. The heavier state is composed mostly of $\psi_L$ and  the orthogonal balanced mixture of $t_R$ and $\psi_R$, whose
mass  $\sim \mu$ supplies the cut-off $\Lambda$ in the effective theory.

\section{Coefficients and Vertex Factors for $\gamma Z$ and $\gamma Z^\prime$}
\label{app:ZA-coeffs}

The coefficients $(C_i)$ for the $Z^\prime \gamma Z$ effective vertex are
given by:
\begin{eqnarray}
C_1 &=& \frac{-2\left(a_t g_V^t + v_t g_A^t \right)}{M_Z^2 - 4 M_\nu^2}
  \biggl[ M_Z^2 \left(B_0(4M_\nu^2;m_t^2,m_t^2) - B_0(M_Z^2;m_t^2,m_t^2)\right) \biggr]
\nonumber\\
  && \,\,\,\,\,\,\,\,\,\,\,\,\,\,\,\,\,\,\,\,\,\,\,\,\,\,\,\,\,\,
     \,\,\,\,\,\,\,\,\,\,\,\,\,\,\,\,\,\,\,\,\,\,\,\,\,\,\,\,\,\,
     \,\,\,\,\,\,\,\,\,\,\,\,\,\,\,\,\,\,\,\,\,\,\,\,\,\,\,\,\,\,
  +2
  \biggl[2m_t^2 \left(3a_t g_V^t - v_t g_A^t\right)
  C_0 + a_t g_V^t + v_t g_A^t \biggr]\,\\
\nonumber\\
C_2 &=& \frac{-2\left(a_t g_V^t + v_t g_A^t \right)}{M_Z^2 - 4 M_\nu^2}
  \biggl[ \left(4M_\nu^2 + M_Z^2 \right)\left(B_0(4M_\nu^2;m_t^2,m_t^2) -
  B_0(M_Z^2;m_t^2,m_t^2)\right)
\nonumber\\
  && \,\,\,\,\,\,\,\,\,\,\,\,\,\,\,\,\,\,\,\,\,\,\,\,\,\,\,\,\,\,
     \,\,\,\,\,\,\,\,\,\,\,\,\,\,\,\,\,\,\,\,\,\,\,\,\,\,\,\,\,\,
     \,\,\,\,\,\,\,\,\,\,\,\,\,\,\,\,\,\,\,\,\,\,\,\,\,\,\,\,\,\,
  +  2 \left(4M_\nu^2 - M_Z^2\right)
  \left(2m_t^2 C_0 + 1\right)\biggr]\,\\
\nonumber\\
C_3 &=& \frac{-4\left(a_t g_V^t + v_t g_A^t \right)}{\left(M_Z^2 - 4 M_\nu^2\right)^2}
  \biggl[ -M_Z^2\left(B_0(4M_\nu^2;m_t^2,m_t^2) - B_0(M_Z^2;m_t^2,m_t^2)\right)
    \nonumber\\
   && \,\,\,\,\,\,\,\,\,\,\,\,\,\,\,\,\,\,\,\,\,\,\,\,\,\,\,\,\,\,
     \,\,\,\,\,\,\,\,\,\,\,\,\,\,\,\,\,\,\,\,\,\,\,\,\,\,\,\,\,\,
     \,\,\,\,\,\,\,\,\,\,\,\,\,\,\,\,\,\,\,\,\,\,\,\,\,\,\,\,\,\,
  - \left(4M_\nu^2 - M_Z^2\right)\left(2m_t^2 C_0 + 1\right)\biggr]\,\\
\nonumber\\
C_4 &=& \frac{4\left(a_t g_V^t + v_t g_A^t \right)}{M_Z^2 - 4 M_\nu^2}
  \left(B_0(4M_\nu^2;m_t^2,m_t^2) - B_0(M_Z^2;m_t^2,m_t^2)\right),\\
\nonumber\\
C_5 &=& \frac{4\left(a_t g_V^t + v_t g_A^t \right)}{\left(M_Z^2 - 4 M_\nu^2\right)^2}
  \biggl[ -M_Z^2\left(B_0(4M_\nu^2;m_t^2,m_t^2) - B_0(M_Z^2;m_t^2,m_t^2)\right)
   \nonumber\\
   && \,\,\,\,\,\,\,\,\,\,\,\,\,\,\,\,\,\,\,\,\,\,\,\,\,\,\,\,\,\,
     \,\,\,\,\,\,\,\,\,\,\,\,\,\,\,\,\,\,\,\,\,\,\,\,\,\,\,\,\,\,
     \,\,\,\,\,\,\,\,\,\,\,\,\,\,\,\,\,\,\,\,\,\,\,\,\,\,\,\,\,\,
  - \left(4M_\nu^2 - M_Z^2\right)\left(2m_t^2 C_0 + 1\right)\biggr]\,,
\end{eqnarray}
where $v_t$ and $a_t$ are the SM vector and axial-vector couplings of a top
quark to a $Z$ boson, i.e.:
\begin{eqnarray}
v_t &=& -1 + \frac{8}{3}s_W^2\,\\
a_t &=& 1 \,.
\end{eqnarray}
and, for the case where the $Z^\prime$ only couples to $t_R$, $g_V^t$ and $g_A^t$ are given
by:
\begin{equation}
g_V^t = - g_A^t = \frac{g_t^{Z^\prime}}{2} \,.
\end{equation}
The corresponding coefficients for the $Z^\prime \gamma Z^\prime$ effective vertex
can easily be obtained by replacing $a_t \to g_A^t$ and $v_t \to g_V^t$ in the 
above expressions.

Finally, the contribution to the $\gamma Z$ matrix-element-squared from the vertex 
factor ${\cal V}_{\gamma Z}^2$ is given in terms of the above $C_i$ coefficients
by:
\begin{eqnarray} 
{\cal V}_{\gamma Z}^2 &=& |C_1|^2 \biggl(\frac{M_Z^2}{4} - 2M_\nu^2 + 
  \frac{4M_\nu^4}{M_Z^2} \biggr) + M_Z^2 |C_2|^2 + |C_5|^2 \biggl(
  \frac{M_Z^6}{16} - M_\nu^2 M_Z^4 + 6 M_\nu^4 M_Z^2 - 16 M_\nu^6 + 16 \frac{M_\nu^8}{M_Z^2}
  \biggr) \nonumber\\
\nonumber\\
&& + (C_1 C_2^* + C_1^* C_2)\left(-\frac{M_Z^2}{2} + 2M_\nu^2\right) + 
  (C_1 C_5^* + C_1^* C_5) \left(-\frac{M_Z^4}{8} + \frac{3}{2}M_\nu^2 M_Z^2 - 
  6M_\nu^4 + \frac{8M_\nu^6}{M_Z^2}\right) \nonumber\\
\nonumber\\
&& + (C_2 C_5^* + C_2^* C_5)\left(\frac{M_Z^4}{4} - 2M_\nu^2 M_Z^2 + 4 M_\nu^2\right)\,,
\end{eqnarray}
where the corresponding factor for $\gamma Z^\prime$ can be obtained from the 
above with the replacement $M_Z \to M_{Z^\prime}$.

\end{document}